\documentclass[amsmath, preprintnumbers, 10pt, twocolumn, pre bibliograpy, floatfix, superscriptaddress]{revtex4-1}

\usepackage{comment}
\usepackage{graphicx}
\usepackage{color}
\usepackage{svg}

\usepackage{physics}
\usepackage{bbm}
\usepackage{braket}
\usepackage[export]{adjustbox}
\usepackage{dsfont}
\usepackage{mathrsfs}
\usepackage{multirow}
\usepackage{float}
\usepackage{tabularx}
\usepackage{caption}
\usepackage[font=small,labelfont=bf,justification=RaggedRight,format=plain, singlelinecheck=false]{caption}
\usepackage{subcaption}
\usepackage{array}
\usepackage{hyperref}
\usepackage{cleveref}
\usepackage{amsmath}
\usepackage{amssymb}
\usepackage{stackengine}
\usepackage{mathrsfs}
\usepackage{ulem}

\renewcommand\theequation{\arabic{equation}}
\DeclareCaptionSubType*{figure}

\renewcommand{\eqref}[1]{Eq.~\ref{#1}}

\usepackage{xr}

\begin{document}
\title{A correspondence between Hebbian unlearning and steady states generated by nonequilibrium dynamics}
\author{Agnish Kumar Behera}
\affiliation{Department of Chemistry, University of Chicago, Chicago, IL, 60637}
\author{Matthew Du}
\affiliation{Department of Chemistry, University of Chicago, Chicago, IL, 60637}
\affiliation{The James Franck Institute, University of Chicago, Chicago, IL, 60637}
\author{Uday Jagadisan}
\affiliation{University of California, Berkeley, CA, 94720}
\author{Srikanth Sastry}
\affiliation{Jawharlal Nehru Center for Advanced Scientific Research, Bengaluru, India}
\author{Madan Rao}
\affiliation{National Center for Biological Sciences, Bengaluru, India}
\author{Suriyanarayanan Vaikuntanathan*}
\affiliation{Department of Chemistry, University of Chicago, Chicago, IL, 60637}
\affiliation{The James Franck Institute, University of Chicago, Chicago, IL, 60637}

\setlength{\abovedisplayskip}{2pt}
\setlength{\belowdisplayskip}{2pt}
\setlength{\abovedisplayshortskip}{2pt}
\setlength{\belowdisplayshortskip}{2pt}

\begin{abstract}

The classic paradigms for learning and memory recall focus on strengths of synaptic couplings and how these can be modulated to encode memories. In a previous paper [A. K. Behera, M. Rao, S. Sastry, and S. Vaikuntanathan,
Physical Review X 13, 041043 (2023)], we demonstrated how a specific non-equilibrium modification of the dynamics of an associative memory system can lead to increase in storage capacity.  In this work, using analytical theory and computational inference schemes, we show that the dynamical steady state accessed is in fact similar to those accessed after the operation of  a classic unsupervised scheme for improving memory recall, Hebbian unlearning or ``dreaming". Together, our work suggests how nonequilibrium dynamics can provide an alternative route for controlling the memory encoding and recall properties of a variety of synthetic (neuromorphic) and biological systems.   

\end{abstract}

\maketitle

\renewcommand*{\thesection}{\Roman{section}}
\section{Introduction}
\renewcommand*{\thesection}{\arabic{section}}

Energy based associative memory models have provided minimal yet powerful frameworks to understand how information storage and retrieval can be modulated in a variety of systems. The Hopfield model and its extensions have found applications for example in Restricted Boltzmann Machines~\cite{marullo2020boltzmann,le2008representational,shimagaki2019selection}, pattern recognition~\cite{schnaack2022learning}, understanding olfaction~\cite{brennan1990olfactory, haberly1989olfactory,wilson2004plasticity} etc. These ideas have also been recently applied to infer models from experimental data for instance data for fitness landscape due to mutations, spike-time correlation data from the brain, etc~\cite{barton2015scaling,cocco2011adaptive,palmer2015predictive}. Moving beyond the original quadratic connectivity, newer variants of the Hopfield model have been explored with higher order connectivity. These have been shown to have exotic properties like exponential capacities~\cite{krotov2023new} and feature-to-prototype transitions in pattern recognition~\cite{krotov2023new,boukacem2024waddington}.

The typical formulation of an associative memory model relies on a local learning strategy, such as the Hebbian rule, for encoding the desired memories or patterns. In a seminal work, Hopfield~\cite{hopfield1983unlearning} showed how the memory capacity of such local associative memory networks maybe improved in an unsupervised and local manner through a so called ``unlearning"  procedure. The unlearning algorithm provides a prescription for updating the connectivity of the associative memory network such that local minima in the free energy landscape which do not correspond to desired patterns are cleaned out leading to a higher capacity. 

Here we consider an alternate paradigm and ask if phemenology resembling Hebbian ``unlearning" (or \textit{dreaming}) can be realized by controlling the dynamics of the spins during pattern retrieval instead of explicitly changing the connectivity beforehand. Using a series of numerical and analytical arguments we demonstrate how such a modulation might be possible. Our central results are numerical and analytical calculations that suggest that the steady states achieved by a particular class of non-equilibrium dynamics resemble steady states achieved when the network is pruned using the Hebbian unlearning rule. We also show how the aforementioned non-equilibrium dynamics may be naturally achievable in a model system of integrate-and-fire neurons. 

The rest of the paper is organized as follows. We begin in Sec.~\ref{Unlearning}, where we review the Hopfield model and the Hebbian unlearning, colloquially referred to as dreaming in Ref~\cite{hopfield1983unlearning}, procedure for improving memory. In the Hebbian unlearning scheme connections between neurons are altered explicitly in an unsupervised manner. In Sec.~\ref{Equivalence} we show the equivalence between dreaming and active dynamics for a continuous version of the Hopfield model. Finally, in Sec.~\ref{DiscreteActivity}, we introduce a version of active dynamics into the standard Hopfield model and numerically demonstrate how activity and Hebbian unlearning have similar qualitative effects on the memory storage and recall properties of the system. We end with Sec.~\ref{Sec:DynamicModulation} and Sec.~\ref{Discussion} where we discuss the implications of our work, biological plausibility, and future directions.

\begin{figure}[tb]
    \centering
    \includegraphics[scale=0.65]{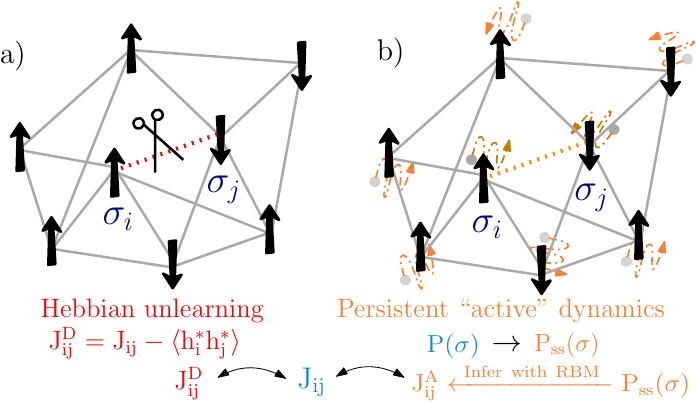}
    \caption{A potential equivalence between Hebbian-unlearning (``Dreaming") and steady states generated during the non-equilibrium dynamics of neurons. (a) The change in synaptic connectivity due to an unsupervised learning scheme, Hebbian unlearning, has been shown to lead to an improvement in the memory capacity. (b) Our analytical and numerical calculations on a model neuronal systems driven by persistent noise sources suggests a systematic connection in which the effective landscapes sampled in the presence of persistent noise sources resembles those generated by Hebbian Unlearning. The synaptic connectivity is inferred from the steady state distribution using an RBM architecture, the details are provided in Sec.~\ref{DiscreteActivity} and Sec.~\ref{RBMNumericalExpt}}
    \label{Schematic}
\end{figure}

\renewcommand*{\thesection}{\Roman{section}}
\section{Hopfield model and Hebbian ``Unlearning"}
\label{Unlearning}
\renewcommand*{\thesection}{\arabic{section}}
The Hopfield model is an associative memory model consisting of $N$ fully connected Ising-like spins, $\sigma_i \in \{\pm 1\}$ for $i=1,\dots,N$, which represent discretized neuron firing rates. The energy of the system is given by the Hamiltonian
\begin{equation}
    H = -\sum_{ij} J_{ij} \sigma_i \sigma_j,
    \label{DiscreteHamiltonian}
\end{equation}
where, following the Hebb rule, the connectivity matrix, $J$ is  
\begin{align}
    J_{ij} = (1/N)\sum_{\mu=1}^{p} \xi^{\mu}_i \xi^{\mu}_j
    \label{eq:genericHebbRule}
\end{align} 
with the patterns $\vec{\xi^{1}},\dots,\vec{\xi^{p}}\in \{\pm 1\}^N$. Using this energy function, the spins can be dynamically updated, synchronously, according to a Metropolis-Monte Carlo algorithm at a chosen temperature scale $T$~\cite{derrida1987exactly}. 
In order to check for retrieval, one initializes the system near a stored pattern (with $\le10\%$ bits flipped) and runs the dynamics for a long time~\cite{hopfield1982neural}. If the final configuration that the system settles in, has a good overlap with the stored pattern (i.e. $\lim_{t\to \infty} \frac{1}{N}\vec{\sigma_{\rm t}} \cdot \vec{\xi^{\rm stored}} \approx 1$), then retrieval is deemed successful. Under such dynamics, the system can store upto $\alpha_c N$ patterns, where $\alpha_c \approx 0.14$ at $T=0$. The retrieval capacity decreases with temperature~\cite{amit1989modeling, amit1985storing}. 

The so called ``Hebbian unlearning" procedure - proposed in Ref.~\cite{hopfield1983unlearning} - and its variants can be used improve the critical memory capacity in an unsupervised manner. The proposed algorithm works as follows. The system is initialized in a random state and allowed to evolve till it reaches a local steady state. Then, the connectivity matrix is updated as,
\begin{align}
    J_{ij}^{D} = J_{ij} - \epsilon \braket{\sigma^*_i \sigma^*_j} \label{HopfieldExplicitDreaming}
\end{align}
where $\vec{\sigma^*}$ is the configuration of the local steady state accessed by the system and $\epsilon$ is the ``unlearning" rate. The system possesses an exponential number, $2^N$, configurations where $N$ is the number of spins of which only a polynomial fraction, $\alpha N$, correspond to stored patterns. If the system starts at a random configuration and performs gradient descent in energy (using \eqref{DiscreteHamiltonian} and \eqref{eq:genericHebbRule}), it has a higher probability of settling into a spurious energy minima than into a pattern basin~\cite{amit1989modeling}. Hence, this unlearning procedure can be viewed as a mechanism to raise the energies of the spurious configurations ($\vec{\sigma^*}$)~\cite{hopfield1983unlearning}. 

\renewcommand*{\thesection}{\Roman{section}}
\section{A plausible equivalence between Hebbian unlearning and active dynamics}
\label{Equivalence}
\renewcommand*{\thesection}{\arabic{section}}
We now argue that there is a plausible equivalence between  Hebbian unlearning and steady states generated by so called \textit{active dynamics} (Fig.~\ref{Schematic}) for a biologically plausible generalization of the Hopfield network, as described by the Hamiltonian 
\begin{equation}
H = -\sum_{ij} J_{ij} f(\sigma_i) f(\sigma_j),
\label{NeuronalActivationHamiltonian}
\end{equation}
where $f$ is a neuronal activation function (e.g., \textit{sigmoid} or $\rm{ReLU}$ function) which mimics the all-or-nothing spiking response of real neurons. The spins $\sigma_i$ are assumed to be continuous variables and evolve according to the dynamics
\begin{align}
    \frac{\partial \sigma_i}{\partial t} =& -\frac{\partial H}{\partial \sigma_{i}} + \eta_i(t), \label{Langevin}
\end{align}
where $\eta_i$ is a Gaussian white noise with statistics
\begin{align}
    \langle \eta_i(t) \rangle = 0 \ , \ \langle \eta_i (t) \eta_j (t') \rangle =& 2T \delta_{ij} \delta(t-t').
    \label{PassiveNoise}
\end{align}

For such generalized Hopfield models, it can be shown that Hebbian unlearning is equivalent to the local learning rule provided in Ref~\cite{dotsenko1991statistical}. In this procedure, a random state $\vec{\sigma}^*$ is chosen and the connectivity matrix is updated as,
\begin{align}
    J_{ij}^{D} = J_{ij} -\epsilon \braket{h^*_i h^*_j},
    \label{eq:DreamingUpdate}
\end{align}
where $h^*_i\equiv -\partial H / \partial \sigma_i = \sum_j J_{ij}f(\sigma_j^*)f'(\sigma_i^*) $ is the effective field felt by spin $i$ when the system is at the random configuration. Plugging $\vec{h}^*$ into the update rule, we find
\begin{align}
     J_{ij}^{D} =& J_{ij} - \epsilon\sum_{p} J_{ip} J_{jp} f'(\sigma_i) f'(\sigma_j) \label{DreamingJ}
\end{align}
Upon many such updates to the connectivity matrix, one can show that the resulting Hamiltonian, with an activation function satisfying $f'(\sigma_i)=1$, takes the form,
\begin{align}
    H = -\frac{1}{N} \sum_{ij \mu \nu} \sigma_i \xi^{\mu}_i (1+\lambda^2 C)^{-1}_{\mu \nu} \xi^{\nu}_j \sigma_j \label{DreamtHamiltonian}
\end{align}
where $\lambda^2=\epsilon t$, $t$ is the time for which the dreaming procedure is carried out. $C$ is the pattern correlation matrix given by, $C_{\mu \nu} = \frac{1}{N} \sum_i \xi^{\mu}_i \xi^{\nu}_i$. Appendix~\ref{Dreaming} details the procedure to obtain \eqref{DreamtHamiltonian} from \eqref{DreamingJ}. If the time, $t$, for dreaming is small, we can express the new connectivity matrix as,
\begin{align}
    J_{ij}^{D} =& J_{ij} - \epsilon t \sum_{p} J_{ip} J_{jp} f'(\sigma_i) f'(\sigma_j) \label{DreamingJ_smalltime}
\end{align}

Next, we show that in a certain perturbative limit, spins when forced out of equilibrium with a persistent noise source (\textit{active dynamics}), reach a steady state which can be expressed using an effective Hamiltonian that has a form similar to that achieved by Hebbian unlearning. Specifically, we consider a noise source with  statistics,
\begin{align}
      \langle \eta_i(t) \rangle = 0 \ , \ \langle \eta_i (t) \eta_j (t') \rangle =& 2T_p \delta_{ij} \delta(t-t') \nonumber \\
    & + \frac{T_a}{\tau} \exp\left(-\frac{|t-t'|}{\tau}\right),
    \label{ActiveNoise}
\end{align}
where $\tau$ is the persistence time of the temporal correlations. Systems with such  persistent dynamics are known to generate non-equilibrium steady states~\cite{fodor2016far}.
For small $\tau$, the spins effectively sample from a distribution given by, $P(\vec{\sigma}) \propto \exp\left( -\beta_{\rm eff} H_{\rm eff} \right)$, where,
\begin{align}
    H_{\rm eff} =& H + \frac{\tau T_a}{T_{\rm eff}} \left( \frac{1}{2}|\nabla H|^2 - T_{\rm eff} \nabla^2 H \right). \label{Heff}
\end{align}
For activation functions like $\rm{ReLU}$ and $\rm{tanh}$, the Laplacian of the Hamiltonian, $\nabla^2H$, has the same sign as the original $J_{ij}$. For instance, when $f(\sigma)=\tanh(\sigma)$, then $f''(\sigma)=-2\sech^2(\sigma)\tanh(\sigma)=-2f(\sigma)\sec^2(\sigma)$ and $J_{ij}^{\rm A} = J_{ij} (1 + \tau T_a\sech^2(\sigma_i)) - \frac{\tau T_a}{T_{\rm eff}}\sum_p J_{ip} J_{jp} |f'(\sigma_p)|^2$. The contribution of the laplacian is a two-body interaction term and is the same sign as the original $J_{ij}$. We expect this to lead to a simple scalar renormalization of the interactions. We ignore this term in the analysis that follows and focus mainly on the $|\nabla H|^2$ term. 

Substituting the form of \eqref{NeuronalActivationHamiltonian} (with only the $|\nabla H|^2$ term) in \eqref{Heff}, we arrive at a renormalized connectivity matrix,
\begin{align}
    J_{ij}^{\rm A} = J_{ij} - \frac{\tau T_a}{T_{\rm eff}}\sum_p J_{ip} J_{jp} |f'(\sigma_p)|^2. \label{ActivityJ}
\end{align}

\begin{figure}[ht]
    \centering
    \includegraphics[scale=0.5]{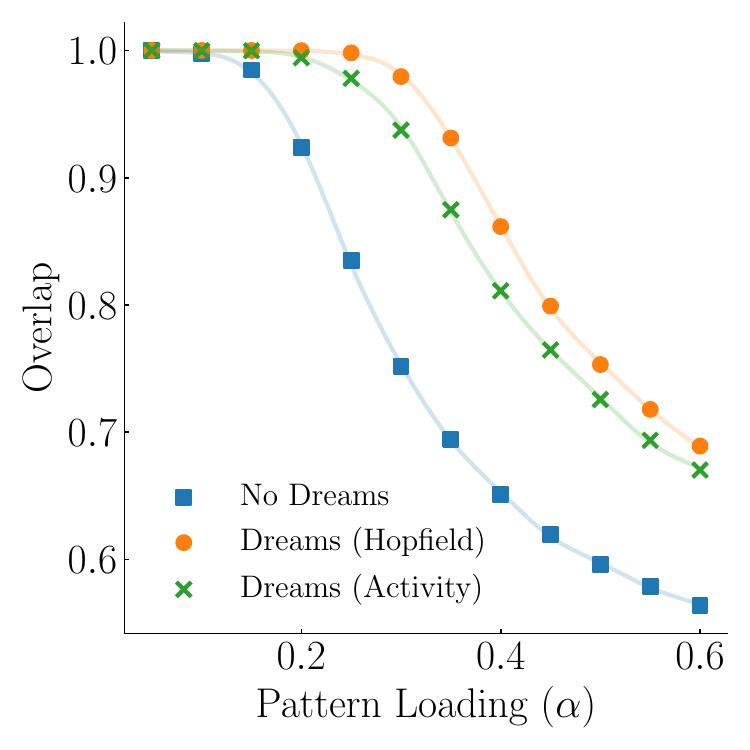}
    \caption{Comparing the information storage limts of \eqref{DreamingJ_smalltime} (Hebbian unlearning) and \eqref{ActivityJ}(Hamiltonian describing the effective steady states under active dynamics). These forms of improving capacity were proposed in Ref.~\cite{dotsenko1991statistical, hopfield1984neurons}. We plot the overlap ($\lim_{t\to \infty} \frac{1}{N}\vec{\sigma_{\rm t}} \cdot \vec{\xi^{\rm stored}}$) averaged over 10 systems and $P=\alpha N$ patterns, for systems of $N=1000$ spins and with interactions specified by \eqref{eq:genericHebbRule}(\textit{Blue}), \eqref{DreamingJ_smalltime}(\textit{Orange}) and \eqref{ActivityJ}(\textit{green}). The $tanh$ activation function was used in all cases. In the absence of any form of dreaming, the overlap decays quickly once the loading around 0.2. For the two dreaming procedures, the construction of the connectivity matrix $J$ is carried out with 10000 dreams and $\epsilon t = \frac{\tau T_a}{T_{\rm eff}} = 10^{-5}$ by repeating \eqref{DreamingJ_smalltime} for the \textit{orange} curve and repeating \eqref{ActivityJ} for the \textit{green} curve. Even with the $tanh$ activation function, \eqref{DreamingJ_smalltime} and \eqref{ActivityJ} have similar qualitative behavior.}
    \label{HopfieldActivityDreaming}
\end{figure}

\eqref{ActivityJ} is  equivalent to \eqref{DreamingJ_smalltime} with the Hebbian unlearning parameter $\lambda^2\equiv \epsilon t = \tau T_a / T_\text{eff}$  and activation function satisfying $f'(\sigma) = 1$.  To numerically verify this connection, we simulate the dynamics of a system of neurons with connectivity given by Eqs.\,\ref{DreamingJ} and \ref{ActivityJ} at zero temperature and compute the ability of the system to retrieve patterns as a function of the number of patterns stored. Both models show the same qualitative behavior as shown in Fig.~\ref{HopfieldActivityDreaming}. This equivalence suggests that the sampling generated by activity in the limit of small persistence times can be similar to that generated by an  ``infinitesimal" Hebbian unlearning procedure. Can this phenomenology persist for arbitrary active dynamics ? Motivated by the equivalence of connectivity from Hebbian unlearning and that due to non-equlibrium activity in the aforementioned perturbative limit, we now use numerical inference schemes to search for a broader equivalence.

\renewcommand*{\thesection}{\Roman{section}}
\section{Numerical inference schemes reveal signatures of Hebbian unlearning in active dynamics}
\label{DiscreteActivity}
\renewcommand*{\thesection}{\arabic{section}}
The results in the previous section suggest that active dynamics driven by persistence noise sources can improve memory in a manner similar to Hebbian unlearning. We now use numerical inference schemes, assisted by Restricted Boltzmann Machine (RBM) architectures, to show that such phenomenology can potentially be observed robustly across many systems. In order to generate configurations that are readily amenable to analysis using an RBM architecture, we switch to spins that can take discrete values. Specifically we start with the standard Hopfield model~\cite{hopfield1983unlearning} and modify the dynamics so that terms reminiscent of persistence are introduced.

\begin{figure}[ht]
    \centering
    \includegraphics[scale=0.5]{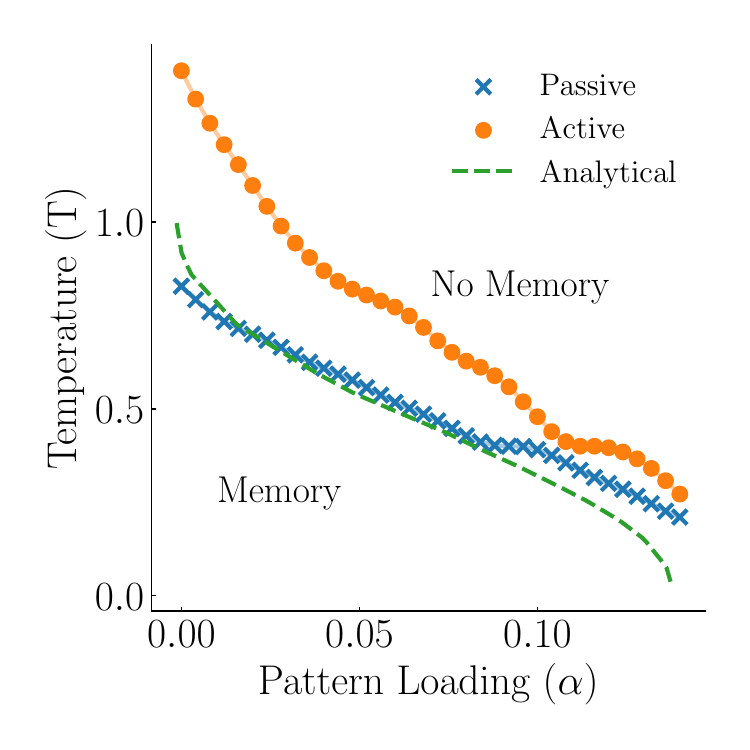}
    \caption{Phase diagram of the discrete Hopfield model under equilibrium spin flips (passive) compared to the phase diagram obtained from the active simulations  in Sec.~\ref{DiscreteActivity}. The plot was made with N=500 spins and averaged over 10 systems and $a=0.5$. The blue `x' denote the region of retrieval for passive dynamics whereas the orange `o' denote the retrieval region for the active (\eqref{NonMarkov}). This result is qualitatively similar to what is observed in the Spherical Hopfield model under activity~\cite{behera2023enhanced}. The analytical curve is taken from Ref.~\cite{amit1989modeling} and is provided as a comparison to the passive numerical simulations.}
    \label{PhaseDiagramComparison}
\end{figure}

To mimic a persistent noise in a discrete system, we designed a process where the evolution of the spins depend on both the current state and the previous state of the system. In particular, the dynamics have an element of persistence. A spin flip (no spin flip) at an instant of time increases the probability of a spin flip (no spin flip) in the subsequent instant of time. 
The evolution of the spins is now Markovian only if both the current and previous state of the system are recorded together. For a single spin, such dynamics can compactly be represented by the following matrix equation
\begin{align}
\begin{pmatrix}
++ \\
+- \\
-+ \\
-- \\
\end{pmatrix}_{t,t+1}
&=\begin{pmatrix}
p+\gamma & 0 & p-\gamma & 0\\
q-\gamma & 0 & q+\gamma & 0\\
0 & p+\gamma & 0 & p-\gamma\\
0 & q-\gamma & 0 & q+\gamma\\
\end{pmatrix}
\begin{pmatrix}
++ \\
+- \\
-+ \\
-- \\
\end{pmatrix}_{t-1,t}
\label{NonMarkov}
\end{align}
where the element $(\sigma \sigma^\prime)_{t,t+1}$ denotes the joint probability of sampling a spin configuration $\sigma$ at time $t$ and $\sigma^\prime$ at time $t+1$. Elements  $(\sigma \sigma^\prime)_{t-1,t}$ have a similar interpretation. 
The typical equilibrium or passive transition probabilities $p_i=1/[1+\rm exp(-2\beta h_i)]$ and $q_i=1-p_i$ are of Boltzmann form, where $\beta = 1/T$ is the inverse temperature and $h_i = \sum_j J_{ij} \sigma_j$ is the effective field felt by spin $i$. To capture the persistence of the continuous active dynamics [\eqref{Langevin} and \eqref{ActiveNoise}], we have introduced the factor $\gamma = a\cdot min(p,q)$, which favours flips to be followed by flips (columns 2 and 3 of transition matrix) and no-flips to be followed by no-flips (columns 1 and 4 of transition matrix). The parameter $a$ controls the persistence, with  $a=0$ corresponding to the passive dynamics. 

Fig.\,\ref{PhaseDiagramComparison} describes the phase diagram of the Hopfield model with the active (persistent) dynamics. For a certain region of the phase diagram, the active dynamics is found to improve pattern retrieval, relative to passive dynamics. This behavior is qualitatively similar to that obtained in previous work with the spherical Hopfield model under activity~\cite{behera2023enhanced}. Note that $T$ here simply refers to the temperature at which the MCMC probabilities of transition, $p$ and $q$, of \eqref{NonMarkov} were calculated. The active dynamics used here does not possess a natural notion of a temperature. 

In order to numerically demonstrate an equivalence similar to \ref{DreamingJ} and \ref{ActivityJ}, we use numerical inference schemes for estimating the connectivity matrices $J_{ij}$ that best explain available data. Extracting the connectivity matrix from samples generated from dynamics is a challenging problem and many techniques have been developed in the context of finding connectivity of real biological neuronal networks~\cite{cocco2011adaptive}. For this study, we used a Restricted Boltzmann Machine (RBM) to extract the connectivity matrix. 

\begin{figure}[ht]
    \centering
    \includegraphics[scale=0.7]{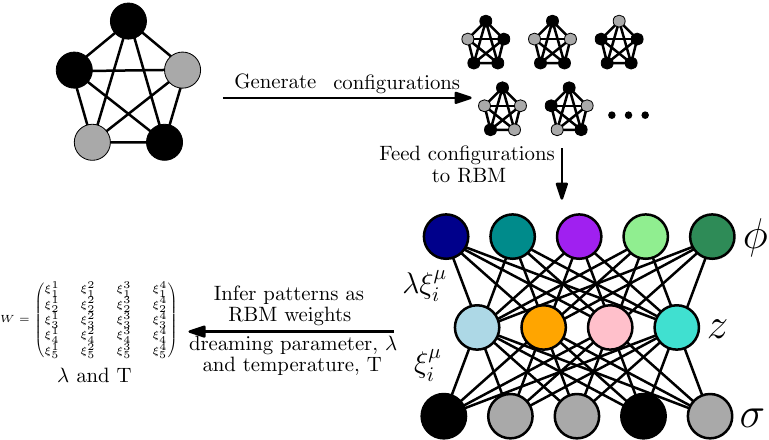}
    \caption{Schematic for the procedure of using RBMs in testing our results. First, we generate a set of configurations at a \textit{specific} temperature using a \textit{specific} dynamics. We denote this temperature by, $T_a$ or $T_p$, where the subscript denotes either passive or active dynamics. Using these configurations, we infer the trainable parameters for the system, RBM weights ($W^{new}$), $T$ and $\lambda$. As described in Appendix~\ref{DifferentModesOfTraining}, we fix the weights of the RBM to the patterns and just learn the dreaming parameter, $\lambda$ and the (``effective") temperature of the simulations.}
    \label{RBMSchematic}
\end{figure}

\begin{figure*}[htb]
    \centering
    \includegraphics[scale=0.9]{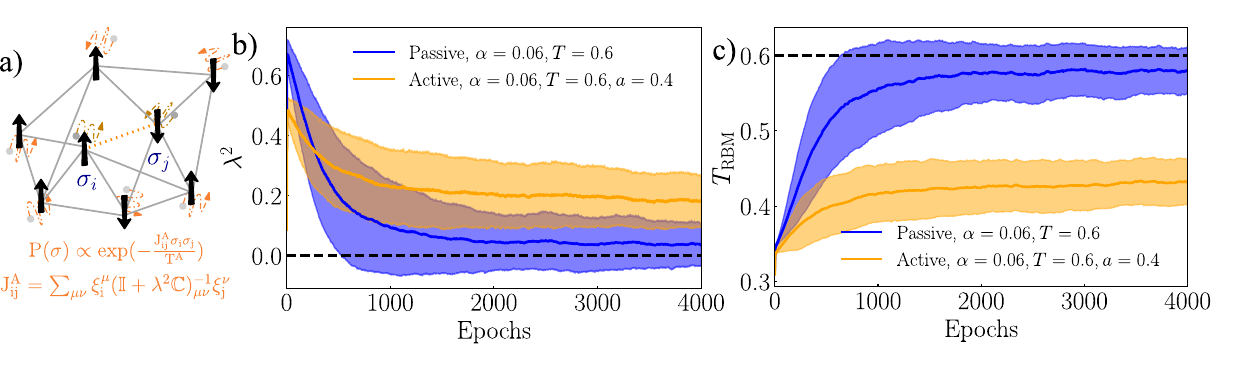}
    \caption{Inferring $T_{\rm RBM}$ and $\lambda$ from data. Panel (a) shows the ansatz for the sampling in the presence of active noise and how the couplings get modified along with the effective temperature. $\lambda=0$ for the passive case where the connectivity becomes the simple Hebbian connectivity. $\lambda>0$ implies ``dreaming" has taken place. In panels (b) and (c), we fix the patterns and let the system learn a $\lambda$ and the temperature for the RBM, $T_{\rm RBM}$. The active case learns a $\lambda$ which is higher than the passive case (which is close to the ground truth of 0). This implies that the sampling due to activity is similar to ``dreaming".}
    \label{ActiveBetterThanPassive}
\end{figure*}

RBMs are a recurrent neural network architecture which have been used extensively for approximating a data distribution and generating further data from the same distribution. RBMs with a single hidden layer have a direct correspondence with the Hopfield model at equilibrium (passive)~\cite{marullo2020boltzmann}. Specifically, if the hidden layer consists of continuous variables with a Gaussian prior, then the probability of observing a certain configuration of hidden and visible nodes is given by,
\begin{align}
    P(\sigma, z) \propto& \ \rm exp\left(-\beta\sum_{\mu}\frac{z_{\mu}^2}{2} - \beta\sum_{\mu i} \frac{\xi^{\mu}_{i}}{\sqrt{N}} z_{\mu} \sigma_i \right)
\end{align}
where $\sigma$ denotes a visible node variable, $z$ a hidden node variable and $\xi^{\mu}_i$ the connection strength between nodes $\sigma_i$ and $z_{\mu}$. To see the connection to Hopfield model one has to marginalize the distribution of configurations by integrating out the hidden variables, $z_{\mu}'s$. This leads to,
\begin{align}
    P(\sigma) \propto& \ \rm 
     \rm exp\left(\beta\sum_{ij} J_{ij} \sigma_i \sigma_j \right) ,\ J_{ij} = \frac{1}{N}\sum_{\mu} \xi^{\mu}_{i} \xi^{\mu}_{j} \label{RBMHebbianEquivalence}
\end{align}
which is equivalent to the equilibrium distribution generated with the Hebbian rule for the Hopfield model. Indeed, the RBM connectivity matrix corresponds to the  patterns variables of the Hopfield model. This makes RBM an ideal system to extract the connectivity matrix.

For our purposes, we need an RBM which can in principle represent the connectivity matrices that can emerge after Hebbian unlearning, the Hamiltonian for which is given by \eqref{DreamtHamiltonian}. This Hamiltonian has two parameters, a parameter $\lambda$ that specifies the extent of Hebbian unlearning and also an effective temperature  $T_{\rm RBM}$. For the passive case, $T_{\rm RBM}$ is equal to the temperature $T$ at which the simulations were carried out. The active simulations lack a notion of ``temperature" and thus $T_{\rm RBM}$ needs to be inferred.  Such a Hamiltonian can be exactly represented as a 3-layer RBM with a visible layer with neurons having discrete values and two hidden layers with continuous-valued neurons~\cite{agliari2019dreaming}. This can readily be seen when the partition function corresponding to the Hebbian unlearned Hamiltonian (\ref{DreamtHamiltonian}) is expressed as a Hubbard-Stratonovich transform,
\begin{align}
    Z =& \int Dz D\phi \sum_{\sigma} \rm exp \left[ -\beta_{\rm RBM} H_{\rm RBM} \right]. 
\end{align} 
where the $H_{\rm RBM}$ is given by 
\begin{align}
    H_{\rm RBM} = \frac{z_{\mu}^2}{2} + \frac{\phi_i^2}{2} - z_{\mu} \frac{\xi^{\mu}_i}{\sqrt{N}}(\sigma_i + i\lambda \phi_i)
\end{align}
and $\beta_{\rm RBM} = \frac{1}{T_{\rm RBM}}$. This expression for the partition function suggests an architecture where a visible layer of neurons is connected to a hidden layer with variables $z_{\mu}$ through weights $\frac{\xi^{\mu}_i}{\sqrt{N}}$ and this hidden layer is again connected to another hidden layer with imaginary variables $\phi_i$ through weights $i\frac{\xi^{\mu}_i\lambda}{\sqrt{N}}$. 
We train this deep hybrid RBM (DHBM) with the data from active and passive simulations and extract the weights $\xi^{\mu}_i$, $T_{\rm RBM}$ and $\lambda$.

We follow a procedure similar to the one outlined in~\cite{leonelli2021effective} to train our RBM with data (see Fig.~\ref{RBMSchematic} for a schematic of the procedure and Sec.\,\ref{RBMNumericalExpt} for additional details).
First, we generate data using the simple Hopfield dynamics (passive) and the non-Markovian dynamics (active). Then we train two different deep hybrid RBMs using this data, one for passive and one for active case.  This training allows us to infer values of $\lambda$ for the two cases. Recall that $\lambda$ is a measure of how much ``dreaming" has taken place. Our simulations (Fig. \ref{ActiveBetterThanPassive}) show that the ``dreaming" parameter" $\lambda$ inferred from the active dynamics indeed has a statistically significant non-zero value. The inference procedure when applied to the passive data leads to values of $\lambda^2$ closer to zero (with some statistical uncertainty). Further, the $T_{\rm RBM}$ learnt for passive simulations is very close to the actual temperature, $T$ at which the data was generated whereas for the active case, $T_{\rm RBM}<T$. Since the reconstruction error is low (see Fig.~\ref{fig:RCEforMainFig}), the Hamiltonian generated by Hebbian unlearning, \eqref{DreamtHamiltonian}, appears to be a good ansatz for the effective Hamiltonian arising from activity. The details of the inference procedure along with additional numerical checks are provided in the Appendix. In Sec.~\ref{subsec:VaryTRBM}, we discuss how the notional ``temperature" can be lower for active simulations.
\begin{figure}[thb]
    \centering
    \includegraphics[scale=0.35]{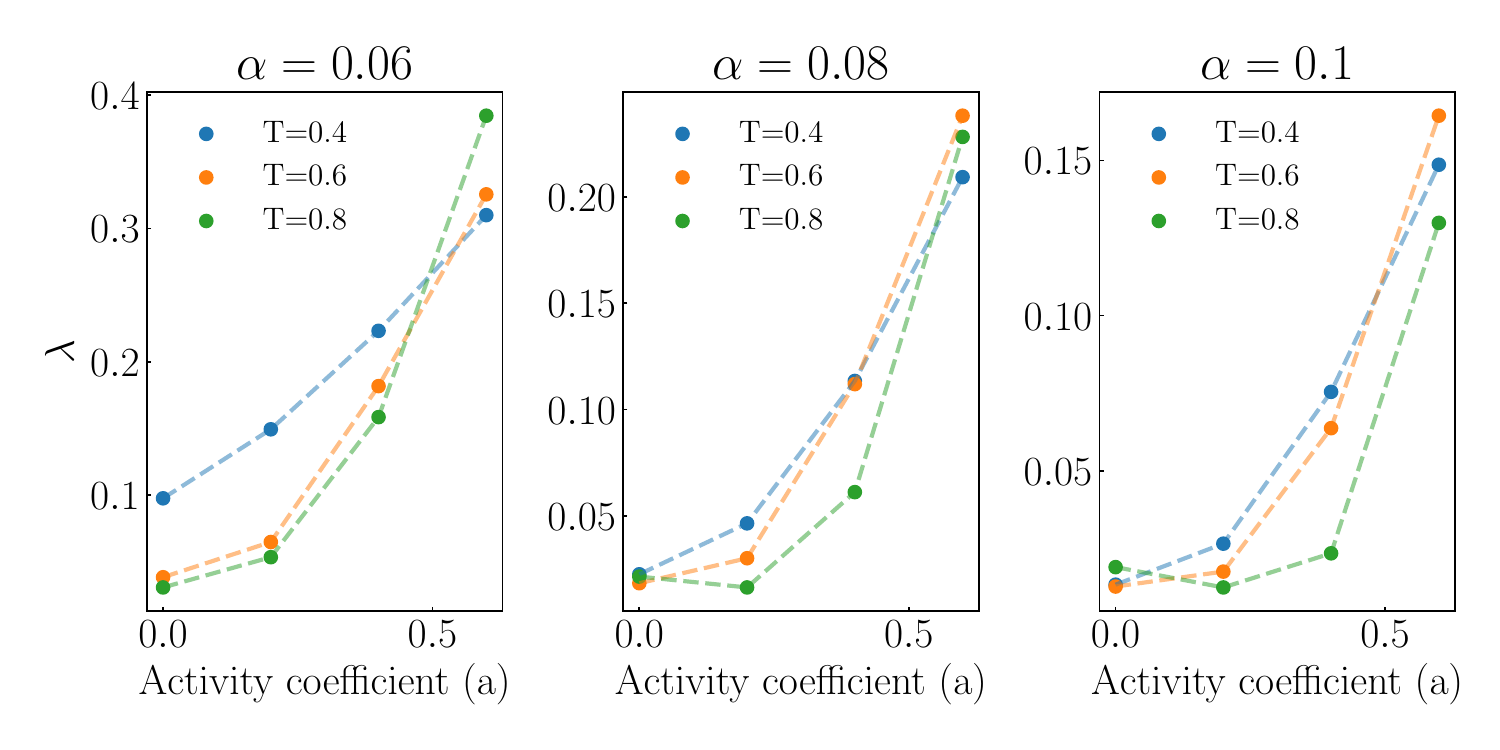}
    \caption{Inferred $\lambda$ values for various combinations of $\alpha$, $T$ and $a$. Increasing $T$, the temperature for data generation does not seem to affect the $\lambda$ significantly. Increasing the non-equilibrium driving by increasing $a$ generally seems to increase $\lambda$ whereas increasing the pattern loading, $\alpha$ seems to decrease $\lambda$.}
    \label{LambdaVsa}
\end{figure}

We also infer $\lambda$ for various combinations of $\alpha$, $T$ and $a$. For the set of inferences in this section, we use mode (3) of Sec.~\ref{appsec:trainingRBM}, i.e. we fix the weights of the RBM to the patterns, $W^{\mu}_i = \xi^{\mu}_i$ and infer $T_{\rm RBM}$ and $\lambda$ values. To ease readability, only the inferred $\lambda$ values have been plotted. In Fig.~\ref{LambdaVsa}, the data was generated using, $\alpha=0.06, 0.08, 0.1$, $T=0.4, 0.6,0.8$ and $a=0.0,0.2,0.4,0.6$ using all combinations. The general trends indicate that $\lambda$ is not sensitive to the temperature, $T$ of data generation but increases (decreases) with increase in $a$($\alpha$). The mild dependence on $T$ is expected because $T$, in the discrete formalism (as given in \eqref{NonMarkov}), is like a simple noise source and should not couple well with the nonequilibrium activity parameter, $a$. Since, $a$ is the parameter for the extent of deviation from equilibrium, the increase of $\lambda$ with $a$ is expected, although one could expect a non-monotonic behavior. The dependence of $\lambda$ on the pattern loading, $\alpha$ is unclear and understanding it is one of our future directions.

\renewcommand*{\thesection}{\Roman{section}}
\section{Dynamical modulation of memory states in neurons: A minimal model}
\label{Sec:DynamicModulation}
\renewcommand*{\thesection}{\arabic{section}}

\begin{figure*}[tbh]
    \centering
    \includegraphics[scale=0.8]{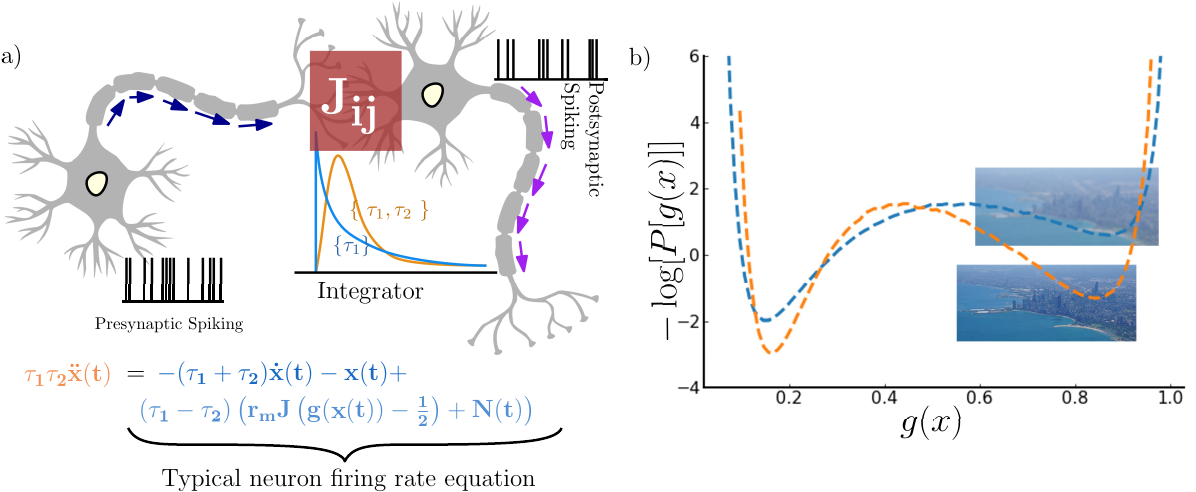}
    \caption{(a)  Temporal dynamics and memory in spiking neurons. The presynaptic signal is integrated at the synapse and helps in the firing of the postsynaptic neuron. The firing rates of the neurons is controlled in important values by the post-synaptic response function $f(t)$ (ESP)
    The blue line denotes the case with just a single post-synaptic decay time constant, $\tau_1\neq 0 \ (\tau_2=0)$ (depicted by the blue integrator) whereas the orange line has two time constants, one for the rising phase and one for decay of the post-synaptic potential, $\tau_1\neq 0, \tau_2\neq 0$ (depicted by the orange integrator). In the equation for firing rate, $g(x)$ denotes an activation function to mimic the all-or-nothing response of neurons.
    (b) Introducing a new time-constant for integration of the potential, leads to a change in the dynamics of firing rates. The resultant steady states can have improved associative memory properties even though the synaptic connections $J_{\rm ij}$ remain unchanged.}
    \label{BialekPlots}
\end{figure*}

While we have specialized most of our results to the case where the extra time scale is introduced in the noise degree of freedom, we anticipate that our findings will be applicable to learning in biological and neuromorphic systems(Fig.~\ref{BialekPlots}). We illustrate this by revealing a connection between a minimal model for neuronal spike generation, adapted from Ref.~\cite{crair1989non}, and the Hopfield model with activity.

To proceed, we first recast the equation of motion for the active Hopfield model [\eqref{Langevin} and \eqref{ActiveNoise}] as underdamped dynamics~\cite{fodor2016far},
\begin{align}
    \tau \ddot{\sigma}_i + \dot{\sigma}_i + \tau \sum_k \frac{\partial^2 H}{\partial \sigma_{i}\partial \sigma_{k}} \dot{\sigma}_k = -\frac{\partial H}{\partial \sigma_{i}} + \eta_i (t), \label{eq:ActiveUnderdamped}
\end{align}
where $\eta_i$ is Gaussian white noise [\eqref{PassiveNoise}]. From this alternative perspective, the persistent motion associated with the time scale $\tau$ can be viewed as a combination of an inertial component [first term on left-hand side (lhs)] and additional damping (third term on lhs), where the latter depends on the firing rates through the activation function, $f$ [\eqref{NeuronalActivationHamiltonian}].

As we next show, analogous dynamical contributions arise from persistence dynamics in a minimal model of neurons that explicitly features spiking.
In this model, which is adapted from Ref.~\cite{crair1989non}, an integrate-and-fire neuron integrates the spiking of neurons in its vicinity and then fires an action potential. The rate of firing of such a neuron, $R_i(t)$ contains information about the state of the entire system and is a Poisson-like process which can be mathematically expressed as,
\begin{align}
    r_i(t) &= r_m g\left(\sum_{j} J_{ij} f\circ R_j(t) - \theta_i \right) \label{ActivationFiring}
\end{align}
where, $R_i(t)$, the firing rate of the $i^{th}$ neuron and $r_i(t)$ is the mean firing rate of neuron $i$,  $R_j(t) = r_j(t) + \eta_j(t)$. Here $\eta_j$ are the fluctuations in that rate with $\langle \eta_j(t) \rangle = 0$ and $\braket{\eta_i (t) \eta_j(t')} = \delta_{ij} \delta(t-t') r_j(t)$ (See Appendix~\ref{SpikingAppendix} for details).  The function $g$ is an activation function which modulates the all-or-nothing like firing behavior of neurons. Here, it is chosen to be $g(x)={1}/{1+e^{-x}}$. The parameter $r_m$ sets the maximum firing rate for neurons, $\theta_i$ is the threshold for neuron $i$,  and $J_{ij}$ denotes the connection strength between neurons $i$ and $j$. The function $f$ represents the post synaptic response function in real neurons and is used as the convolution window by the post synaptic neuron to integrate the spiking of neurons in it's vicinity. The operation $f\circ R$ denotes a convolution i.e. $(f \circ R)(t) = \int f(t-t') R(t') \Theta(t-t') dt'$ where $\Theta(t-t')$ denotes the Heaviside function and is required to ensure causality. The function $f$ can generically be expected to have a rise and a fall time scale, $f=\exp(-t/\tau_1)-\exp(t/\tau_2)$. In particular, the rise time plays the role of persistence, where past spikes (from times $~\tau_2$ ago) have greater influence on the dynamics, while more recent spikes have less (Fig. \ref{BialekPlots}a). 

As outlined in the Appendix~\ref{SpikingAppendix}, by leveraging the form of $f$ in \eqref{ActivationFiring}, the equations for the firing rates take the form (for a two neuron system),
\begin{align}
    \tau_1 \tau_2 \dot{p(t)} &+  \tau_1 p(t) + \tau_2 p(t) + x(t) = r_mJ \left( y(t) - \frac{1}{2}\right) + N(t) \label{EOMofSpiking_twoneurons}
\end{align}
where $y_i(t) = {r_i(t)}/{r_m}$ and, $x_i(t)\equiv g^{-1}(y_i(t))$ and  $p_i(t) \equiv {d x_i(t)}/{dt}$, $y(t) \equiv \frac{1}{2}(y_1(t) + y_2(t)), x(t) \equiv \frac{1}{2}(x_1(t) + x_2(t)), p(t) \equiv \frac{1}{2}(p_1(t) + p_2(t))$. For large firing rates, we can approximate $N(t)$ to be a Gaussian process, $\braket{N(t)}=0, \braket{N(t)N(t')}=\delta(t-t')y(t)$ ~\cite{crair1989non}.  \eqref{EOMofSpiking_twoneurons} has features similar to those in \eqref{eq:ActiveUnderdamped}. Specifically, the persistent dynamics due to the $\tau_2$ rise time includes inertial (first term of lhs in both equations) and dissipative (third term of lhs in both equations) contributions. 

However, a key difference lies in the dependence (or lack thereof) of the dissipative contribution on the firing rates. Indeed, if the time scales $\tau_1$, $\tau_2$ depend in specific ways on the firing rates themselves, then \eqref{EOMofSpiking_twoneurons} could be mapped, or at least would be more strongly connected, to the equation of motion for the active Hopfield model [\eqref{eq:ActiveUnderdamped}]. Then, given the numerical and analytical results in the previous sections of the paper, it might be possible to argue that a Hebbian unlearning like mechanism can be accomplished dynamically without explicitly tuning the connectivity strength. Can a minimal biophysical model permit the time scales $\tau_{1\rm,2}$ to depend on the firing rates $y$ ? 

The modulation of rise and fall times may naturally occur in a neuron due to the integration of temporally resolved signals by the dendrites~\cite{stuart2016dendrites}. A plausible mechanism is discussed in Sec.~\ref{appsec:IonChannel} and is motivated from  Ref.~\cite{dayan2005theoretical}. In this mechanism, every presynaptic action potential leads to a rise in neurotransmitters which in turn lead to the activation of receptors on the postsynaptic membrane. The probability that the receptors stay activated is denoted as $P_s$. Intuitively, we expect that the integration of spikes through $f$ is a measure of the neurotransmitters effectively taken up by the post-synaptic membrane. In turn, this is a good proxy for the proportion of receptors activated at the post-synaptic membrane. Thus, we expect $P_s$ to be analogous to the integral due to spikes, given by $f\circ R_j(t)$ in \eqref{ActivationFiring}. We can now explore the dependence, if any, of the time scales $\tau_{1,2}$ on the firing rates by studying the dynamics of $P_s$ in various regimes. 

A minimal phenomenological rate equation for $P_s$ can be written as,
\begin{align}
    \frac{dP_s(t)}{dt} &= \alpha_s(r, t) (1-P_s(t)) -\beta_s P_s(t)
\end{align}
where $\alpha_s(r,t)$ denotes the rate of opening and $\beta_s$ denotes the rate of closing of post-synaptic receptors, $r$ is the rate of firing on the pre-synaptic neuron, $r\equiv r_m y$. In  Sec.~\ref{appsec:IonChannel}, we show using a minimal model, how for certain forms of $\alpha_s(r,t)$, $\tau_2$ can vary as a function of $r$. Specifically, in this minimal model (see Sec.~\ref{subsec:Model2}), $\alpha_s(t)$ increases incrementally with every spike (see Fig.~\ref{fig:Model1Model2}(b)) and such a model leads to $\tau_2 \to \tau_2(y)$. Fig.~\ref{fig:Tau2Vsr} shows the dependence of $\tau_2$ on firing rate, $r$.

The $\tau_2(y) p_i$ term in \eqref{EOMofSpiking_dependent} gives us an active matter like flavor where the ``position" degrees of freedom ($y_i$) are now coupled to the ``momentum" degrees of freedom ($p_i$) [\eqref{eq:ActiveUnderdamped}]. Numerical simulations show that the $\tau_2$ rise time increases the stability of the pattern configurations (Fig. \ref{BialekPlots}b), implying a dynamical strengthening of the connectivity.

This minimal mechanism shows how it might be possible for neuronal systems to dynamically change their time scales and \textendash given the results of the previous sections \textendash mimic Hebbian unlearning like processes without explicitly changing the synaptic strengths. 

\renewcommand*{\thesection}{\Roman{section}}
\section{Discussion}
\label{Discussion}
\renewcommand*{\thesection}{\arabic{section}}
The information storage and processing ability of neurons is constantly modified by their inherent synaptic plasticity.  The interplay between synaptic dynamics and the memory storage and retrieval properties of neurons has broadly been well studied in many seminal works ~\cite{amit1985storing}. Our work here suggests how phenomenology resembling synaptic plasticity can be acheived by simply introducing an extra time scale in the dynamics of the neuron like degrees of freedom. 

The main analytical and numerical work presented in this paper considers a model system in which an extra time scale is introduced in the dynamics of the neuronal degrees of freedom. This extra time scale is introduced as a persistence or correlation time in the noise driving the neuronal degrees of freedom. In previous work~\cite{behera2023enhanced} we had demonstrated how such persistent or \textbf{active} degrees of freedom can potentially lead to better memory recall properties. In this work, we have demonstrated that such persistent degrees of freedom create an information storage landscape resembling one created by an unsupervised learning strategy commonly referred to as Hebbian unlearning or dreaming~\cite{hopfield1984neurons}. 

 In the simple model in Fig.~\ref{BialekPlots}, the extra time scale is modulated mainly by the dynamics of the postsynaptic response function, $f$. This function mimics the receptor opening and closing probability at the post-synaptic membrane and can be influenced by a variety of factors, firing rate being one of them~\ref{appsec:IonChannel}. Other modes for introducing extra time scales include the coupling between neuronal and transcriptional dynamics, the spatial summation of  post-synaptic potentials due to multiple neurons, the topology of the connections etc.~\cite{stuart2016dendrites}.  These might also lead to an improvement in the memory recall properties.  \\

Finally, we anticipate that our findings maybe applicable in other contexts  such as protein and chromatin folding problems which can be posed as associative memory problems~\cite{zheng2012predictive, beissinger1998chaperones}. Due to the frustration implicit in these systems the task of finding the desired ground structural state is highly non-trivial. Our work suggests how adding extra time scales in the relaxation process, these could for example be possible due to the action of molecular chaperons that are known to consume ATP, may help sculpt the landscape and aid in the process of finding the desired or programmed state~\cite{goloubinoff2018chaperones,saibil2013chaperone}. Together, our work simply provides a framework for understanding how non-equilibrium relaxational dynamics with multiple time scales $\textendash$ these occur routinely in biology $\textendash$ can be used to improve the effectiveness of memory recall.

\renewcommand*{\thesection}{\Roman{section}}
\section{Acknowledgments}
\label{Acknowledgments}
\renewcommand*{\thesection}{\arabic{section}}
This work was mainly supported by funds from DOE BES Grant No. DE-SC0019765 to S.V. and M.D.  A.K.B. acknowledges support from a fellowship from the Department of Chemistry at the University of Chicago. We also acknowledge support from the National Science Foundation through the Physics Frontier Center for Living Systems (PHY-2317138). We acknowledge helpful discussions with Brent Doiron.

\setcounter{figure}{0}
\setcounter{table}{0}
\setcounter{equation}{0}
\setcounter{page}{1}
\setcounter{section}{0}

\renewcommand{\theequation}{A\arabic{equation}} 
\renewcommand{\thepage}{A\arabic{page}} 
\renewcommand{\thesection}{A\arabic{section}} 
\renewcommand{\thesubsection}{A\arabic{section}.\arabic{subsection}}
\renewcommand{\thetable}{A\arabic{table}}  
\renewcommand{\thefigure}{A\arabic{figure}}

\section{Integrate-and-Fire model}
\label{SpikingAppendix}

Let us first define the integrators for the case with a single exponential fall phase ($f^{(1)}$) and the one with a rise and fall phase ($f^{(2)}$) and the corresponding operators for those integrators ($\mathcal{O}^{(1)}$ and $\mathcal{O}^{(2)}$).
\begin{align}
    f^{(1)}(t) &= \frac{1}{\tau_1} \exp(-\frac{t}{\tau_1}) \\
    f^{(2)}(t) &= \frac{1}{\tau_1 - \tau_2} \left( \exp(-\frac{t}{\tau_1}) - \exp(-\frac{t}{\tau_2}) \right) \\ 
    \mathcal{O}^{(1)} &= \left( \frac{d}{dt} + \frac{1}{\tau_1} \right) \\
    \mathcal{O}^{(2)} &= \left( \frac{d}{dt} + \frac{1}{\tau_1} \right)\left( \frac{d}{dt} + \frac{1}{\tau_2} \right)
\end{align}

From \eqref{ActivationFiring} we can derive the equation of motion for the rate of firing of the neurons. We will write down the equations of motions for both the single exponential and the rise and fall integrator.

\begin{align}
    x_i &+ \theta_i =  \sum_{j} J_{ij} f^{(1,2)}\circ (r_m y_j + \eta_j)\\
    y_i &= \frac{r_i}{r_m}, \ x_i=g^{-1}(y_i), \ p_i =\dot{x_i} = \frac{dx_i}{dt} \\
    \mathcal{O}^{(1)} (x_i &+ \theta_i ) =  \mathcal{O}^{(1)} \sum_{j} J_{ij} f^{(1)}\circ (r_m y_j + \eta_j) \\
    \tau_1 \dot{x_i} &= -x_i -\theta_i + r_m\sum_j J_{ij}\left(y_j - \frac{1}{2} \right) + N_i \\
    \mathcal{O}^{(2)} (x_i &+ \theta_i ) =  \mathcal{O}^{(2)} \sum_{j} J_{ij} f^{(2)}\circ (r_m y_j + \eta_j) \\
    \tau_1 \tau_2 \dot{p_i} &+  (\tau_1 + \tau_2)p_i + x_i + \theta_i \nonumber \\
    &= r_m \sum_j J_{ij} y_j + N_i \label{EOMofSpiking_raw} \\
    N_i(t) &= \sum_j J_{ij} \eta_j 
\end{align}
It is important to note that $N_i(t)$ is a state-dependent noise function. We can choose the value of threshold, $\theta_i$ such that \eqref{EOMofSpiking_raw} reduces to,
\begin{align}
    \tau_1 \tau_2 \dot{p_i} &+  (\tau_1 + \tau_2)p_i + x_i = r_m \sum_j J_{ij} \left(y_j - \frac{1}{2}\right) + N_i \label{EOMofSpiking}
\end{align}

\section{Connection to Active matter dynamics}
\label{sec:connActMatt}
The equation of motion for particles undergoing dynamics in the presence of active noise is given as,
\begin{align}
    \dot{x}_i =& -\nabla_i{\phi} + \eta_i \\
    \tau \dot{\eta}_i =& -\eta_i + \xi_i \\
    \braket{\xi_i(t)} =& 0, \ \braket{\xi_i(t) \xi_j(t')} = 2T\delta_{ij} \delta(t-t')
\end{align}
where, $\phi$ is the potential energy, $\tau$ is the time-scale of active fluctuations $\eta$ and $\xi$ is standard white noise. This leads to the noise, $\eta$ have correlations of the form, 
\begin{align}
    \braket{\eta_i(t)} = 0 \, \ \braket{\eta_i(t) \eta_j(t)} = \frac{T}{\tau} \exp(-\frac{|t-t'|}{\tau})
\end{align}

Denoting $\dot{x} = p$, we can rewrite this set of equations as~\cite{fodor2016far},
\begin{align}
    \tau \dot{p}_i = - p_i + (1 + \tau p_k\cdot \nabla_k) \nabla_i{\phi} + \eta \label{eq:ActMatEqn}
\end{align}
We will now try to connect \eqref{eq:ActMatEqn} to \eqref{EOMofSpiking}. The connection between the two equations is the introduction of the rise time scale $\tau_2$ in \eqref{EOMofSpiking}. Specifically, when analyzed around one of the stable firing rates, Eq.~\ref{EOMofSpiking} has an extra frictional component $\tau_2 p_i(t)$ similar to that encountered for a particle trapped in a harmonic well and driven by noise with a persistent degree of freedom (Fig.~\ref{BialekPlots}(b)). Note that this analysis and analogy is limited to cases where $y(t)$ is fluctuating around one its metastable points. In general, the noise function $N(t)$ is state dependent and an exact analogy is no longer possible. 

The main models used in this work share some of the broad qualitative features of Eq.~\ref{EOMofSpiking} and active matter models (\eqref{eq:ActMatEqn}). The dynamics of the discrete Hopfield like model in Section~\ref{DiscreteActivity} takes into account spin flips in the immediate history qualitatively bringing in a new time-scale into the picture. In order to have a more concrete connection to active matter, we need to have a coupling term like $p_k \cdot \nabla_k \nabla_i{\phi}$ as in \eqref{eq:ActMatEqn}. This is possible in \eqref{EOMofSpiking} if $\tau_1$ and $\tau_2$ are functions of the rate of spiking, $y$. In the next section, we discuss how this is achieved in a biological setting.

\section{Receptor dynamics at the post synaptic membrane and a time scale dependent on rate of firing}
\label{appsec:IonChannel}
Synaptic transmission occurs when a spike arrives at the presynaptic terminal and leads to the release of neurotransmitters inside the synaptic cleft (the region between the two neurons at the synapse). These neurotransmitters are then absorbed by the postsynaptic neuron through various channel proteins~\cite{dayan2005theoretical}. The synaptic conductance of the current is thus proportional to the product of two terms, $P = P_{\rm rel}P_s$, where $P_{\rm rel}$ is the release probability of neurotransmitters given an action potential arrives at the terminal and $P_s$ is the probability that the postsynaptic channel opens given neurotransmitters were released at the cleft. We assume a constant $P_{\rm rel}$ and focus mainly on $P_s$ in what follows. $P_s$ can be modelled as a directly activated receptor channel where the transmitter binds directly with the channel to open it and then unbinds after a certain amount of time to close it~\cite{dayan2005theoretical}. The following equation can be used to describe $P_s$,
\begin{align}
    \frac{dP_s(t)}{dt} = \alpha_s(r, t) (1-P_s(t)) - \beta_s P_s(t) \label{eq:Ps}
\end{align}
Here, $\beta_s$ is the closing rate of the channel and is usually assumed to be a constant. The opening rate, $\alpha_s(r,t)$ depends on the concentration of the transmitters, which changes with the spiking rate, $r$, and time, $t$.

Under certain conditions, we can express $P_s$ in terms of the convolution kernel for integration, $f(t)$ in Sec.~\ref{SpikingAppendix}. To see this, consider \eqref{eq:Ps} with $\alpha_s(t) = P_0 \sum_j \delta(t-t^j)$, where $t^j$ is the time of spike $j$, and $\beta_s$ constant. In this limit, after every spike, we can write, $P_s \to P_s + P_0$, and between spikes, $P_s(t) = P_s(0)e^{-\beta t}$. Thus, under the assumption that $P_0\ll1$, the functional form of $P_s$ after multiple spikes is given as, $P_s(t) = P_0\sum_{j} e^{-\beta (t - t^{j})}$. This functional form is nothing but the convolution of the spike train with the kernel, $f=e^{-\frac{t}{\tau}}$ with $\beta=\frac{1}{\tau}$.

In the next subsection, Sec. \ref{subsec:Model1} we describe ``Model 1'' where we similarly show, for appropriate choice of $\alpha_s(r,t)$, that $P_s$ can be approximated by the convolution of the spike train $R(t)$ with the kernel, $f=e^{-\frac{t}{\tau_1}} - e^{-\frac{t}{\tau_2}}$. In Sec.~\ref{subsec:Model2}, we discuss a different model, ``Model 2'', having a different functional form for $\alpha_s(r,t)$. For both the models, we calculate the steady state probabilities, $P_s^{(ss)}$ and equate them. By doing so, we find an effective description of Model 2 in terms of Model 1 parameters. Then we can use Model 1 with these effective parameters and connect it to the convolution kernel, $f=K(e^{-\frac{t}{\tau_1}} - e^{-\frac{t}{\tau_2}})$. For further simplification, we also assume that the spikes are uniformly distributed in time, i.e., $t^{j} = j / r$ for firing rate $r$.

\subsection{Model 1}
\label{subsec:Model1}
\begin{figure}[htb]
    \centering
    \includegraphics[width=1.0\linewidth]{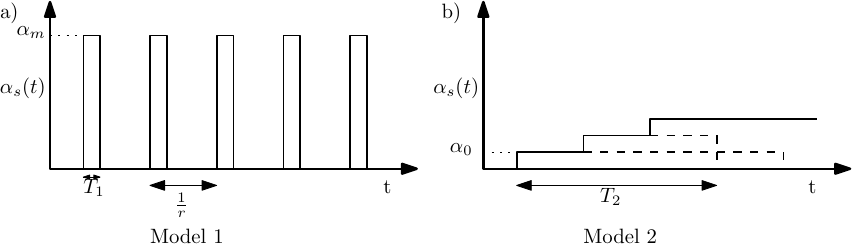}
    \caption{Schematic for $\alpha_s(t)$ profiles for Model 1 and Model 2. The dashed lines in panel (b) denote the course of $\alpha_s(t)$ due to the first and second spikes.}
    \label{fig:Model1Model2}
\end{figure}

In ``Model 1'', after ever presynaptic spike, the postsynaptic receptors display a very fast Markovian behavior of switching between ``open'' and ``closed'' states.This can be expressed as an instantaneous rise in $\alpha_s(t)$ to its maximum value, $\alpha_m$. $\alpha_s(t)$ remains constant at that value for a specific duration, $T_1$ before instantaneously dropping to 0~\cite{dayan2005theoretical}. Thus $\alpha_s(t)$ is a rectangular pulse function. Mathematically, it can be expressed as, 
\begin{align}
    \alpha_s(t) = \mathbb{T}(\alpha_m\sum_{f} [\theta(t-t^f) - \theta(t-t^f-T_1)] , \alpha_m)  \label{eq:alpha_sModel1}
\end{align} 
where, $\theta$ i s the Heaviside function, $t^f$ are the spiking times and  $\mathbb{T}(a,b)$ is the threshold function which ensures that the function does not exceed $b$. A schematic of this form of $\alpha_s(t)$ is plotted in Fig.~\ref{fig:Model1Model2}(a). This generates a $P_s$ profile for Model 1 for which an excellent approximation is $f\circ R$ with the convolution kernel, $f=K(e^{-\frac{t}{\tau_1}} - e^{-\frac{t}{\tau_2}})$, when a single spike is fired. The corresponding $\tau_1$ and $\tau_2$ are given by, $\tau_1 = \frac{1}{\beta}$ and $\tau_2 \sim \frac{1}{\alpha_m + \beta}$. The exact functional form of $\tau_2$ is pretty complicated but it roughly goes as the inverse of the rate of rise of $P_s$ i.e. $\alpha_m$.

Using \eqref{eq:alpha_sModel1}, we can calculate the steady state value of $P_s$. Let's say $P_s$ was $P_s^{(ss)}$ just before the arrival of a presynaptic spike. After the spike arrives, it rises to a value, $P_s^{max}$ in time $T_1$ and then decays for a time, $\frac{1}{r} - T_1$ (uniform spiking with rate r leads to interspike intervals of length $\frac{1}{r}$) before the arrival of a new spike. For this model, $T_1<<\frac{1}{r}$. This limit is to ensure that $\alpha_s(t)$ decays to $0$ between two consecutive spikes. We want the rise in $P_s$ and its decay to compensate for one another.
\begin{align}
    \frac{dP_s}{dt} =& \alpha_m(1-P_s) - \beta_s \ \rm for \ 0<t<T_1 \\
    P_s(t=T) =& P_s^{(ss)} e^{-(\alpha_m + \beta)T_1} + \frac{\alpha_m}{\alpha_m + \beta} \left[ 1 - e^{-(\alpha_m + \beta)T_1} \right] \\
    \frac{dP_s}{dt} =& -\beta P_s \ \rm for \ T_1<t<\frac{1}{r}\\
    P_s\left(t=\frac{1}{r}\right) =& P_s(t=T) e^{-\beta(\frac{1}{r} - T_1)} = P_s^{(ss)} \\
    \implies P_s^{(ss,1)} =& \frac{\alpha_m}{\alpha_m + \beta} \frac{e^{\beta T_1} - e^{-\alpha_m T_1}}{e^{\beta/r} - e^{-\alpha T_1}} \label{eq:SimpleConv}
\end{align}

\subsection{Model 2}
\label{subsec:Model2}
Now, we will discuss ``Model 2''. In this model, the post synaptic receptors display a cooperative behavior with every presynaptic spike aiding in opening more receptors in the post synaptic membrane. Every presynaptic spike opens a fraction $\alpha_0$ of postsynaptic receptors and the time scale of this cooperative process is $T_2$. In Model 2, $T_2>\frac{1}{r}$. This limit is to ensure that cooperativity persists for longer than the interspike interval. Mathematically, the $\alpha_s$ function and its steady state value of $\alpha_s$ can be expressed as,
\begin{align}
    \alpha_s(r, t) =& \alpha_0 \sum_{f=1}^n \left[ \theta(t-t^f) - \theta(t-t^f-T_2) \right] \\
    \alpha_s^{(ss)} =& \alpha_0 rT_2
\end{align}
A schematic of $\alpha_s(t)$ for Model 2 is plotted in Fig.~\ref{fig:Model1Model2}(b). With this form of $\alpha_s$, the final steady state probability $P_s^{(ss),2}$ is given by,
\begin{align}
    P_s^{(ss,2)} = \frac{\alpha_0 rT_2}{\alpha_0 rT_2 + \beta}
    \label{eq:ComplexConv}
\end{align}

\subsection{Equating the steady state probabilities}
\label{subsec:EquatePs}
\begin{figure}[htb]
    \centering
    \includegraphics[width=0.8\linewidth]{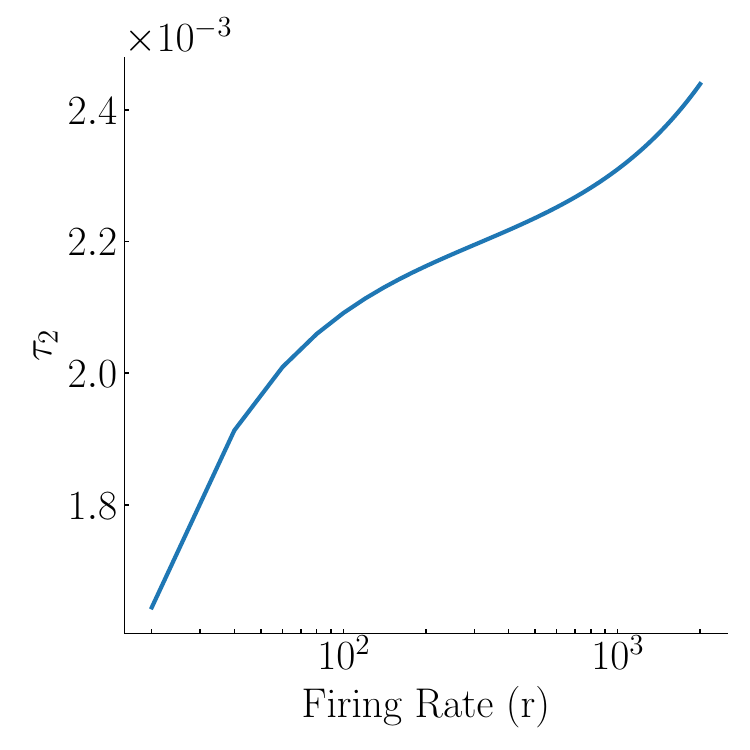}
    \caption{$\tau_2$ increases as the rate of firing increases. With the increase in the rate of firing the steady-state probability, $P_s^{(ss)}$ increases. In order to have a late onset of decay, the rise time, $\tau_2$, has to increase. The parameters used for this simulation: $\alpha_0=2, T_2=0.1,T_1=5\times10^{-4},\beta=10$.}
    \label{fig:Tau2Vsr}
\end{figure}

Now we equate the steady state probabilities, $P_s^{(ss)}$ from both the models in \eqref{eq:SimpleConv} and \eqref{eq:ComplexConv}. We do it since we want a description of Model 2 in terms of the parameters of Model 1, i.e. we want an ``effective'' $\alpha_m$ and $\beta$ for ``Model 2''. The procedure to extract these ``effective'' parameters is as follows.
\begin{align}
    \frac{\alpha_m}{\alpha_m + \beta} \frac{e^{\beta T_1} - e^{-\alpha_m T_1}}{e^{\beta/r} - e^{-\alpha T_1}} = \frac{\alpha_0 rT_2}{\alpha_0 rT_2 + \beta} \label{eq:SimpleComplexEquality}
\end{align}
We assume that the rate of decay, $\beta$ is the same in both models and we also fix the duration, $T_1$ and $T_2$. Then we solve for $\alpha_m$ in \eqref{eq:SimpleComplexEquality}. This is a transcendental equation and we use numerical tools to compute $\alpha_m$. Now we use these effective parameters to obtain $\tau_1$ and $\tau_2$ for the convolution kernel, $f$. We can do so because Model 1 is equivalent to $f$ for a single spike. After solving for $\alpha_m$, we use this value of $\alpha_m$, $\beta$, $r$ and $T_1$ to construct a new $P_s(t)$ profile for a single spike. Now, we fit this profile to the function, $f(t) = B\left( \exp(-\frac{t}{\tau_1}) - \exp(-\frac{t}{\tau_2}) \right)$ and extract $\tau_1$ and $\tau_2$. We repeat this procedure keeping the variables, $\beta$, $T_1$, $T_2$ and $\alpha_0$ fixed and varying only the rate of firing $r$. Fig.\ref{fig:Tau2Vsr} shows how $\tau_2$ varies as a function of the firing rate $r$ in this equivalent description.
This $\tau_2(y)$ can be replaced in \eqref{EOMofSpiking} and that leads to,
\begin{align}
    \tau_1 \tau_2(y) \dot{p_i} &+  (\tau_1 + \tau_2(y))p_i + x_i = r_m \sum_j J_{ij} \left(y_j - \frac{1}{2}\right) + N_i \label{EOMofSpiking_dependent}
\end{align}
The $\tau_2(y) p_i$ term in \eqref{EOMofSpiking_dependent} give us an active matter like flavor where the ``position" degrees of freedom ($y_i$) are now coupled to the ``momentum" degrees of freedom ($p_i$) which in turn can give rise to novel phenomena.

 \section{Dreaming}
\label{Dreaming}
The evolution of the connectivity matrix following the procedure of \eqref{eq:DreamingUpdate} can be expressed mathematically as,
\begin{align}
    J_{ij}(t+1) = J_{ij}(t) - \epsilon \langle h_i^*(t) h_j^*(t) \rangle
\end{align}
For the Hamiltonian in \eqref{NeuronalActivationHamiltonian}, the process of dreaming can be mathematically expressed as,
\begin{align}
    J_{ij}(t+1) =& J_{ij}(t) - \epsilon\sum_{pq} \langle J_{ip} J_{jq} f(\sigma_p) f(\sigma_q) f'(\sigma_i) f'(\sigma_j) \rangle \\
    =& J_{ij}(t) - \epsilon\sum_{pq} J_{ip} J_{jq} \langle f(\sigma_p) f(\sigma_q) f'(\sigma_i) f'(\sigma_j) \rangle \\
    =& J_{ij}(t) -  \epsilon\sum_{pq} J_{ip} J_{jq} \delta_{pq} f'(\sigma_i) f'(\sigma_j) \\
    =& J_{ij}(t) - \epsilon\sum_{p} J_{ip} J_{jp} f'(\sigma_i) f'(\sigma_j) \label{DreamingJ_2}
\end{align}
For $f'(\sigma) = 1$, \eqref{DreamingJ} can be recast as,
\begin{align}
    \frac{\delta J}{\delta t} =& - \epsilon J^2 \\
    \implies J(t) =& \frac{J(0)}{(1 + \epsilon t J(0))}, \ J(0)_{ij} = \frac{1}{N} \sum_{\mu} \xi^{\mu}_i \xi^{\mu}_j \\
    J_{ij} =& \frac{1}{N} \sum_{\mu \nu} \xi^{\mu}_i (1 + \lambda^2 C)^{-1}_{\mu \nu} \xi^{\nu}_j \label{JijofDreamtHamiltonian}\\
    C_{\mu \nu} =& \frac{1}{N} \sum_i \xi^{\mu}_i \xi^{\nu}_i \ , \ \lambda^2= \epsilon t \label{ExtentOfDreaming}
\end{align}
\eqref{JijofDreamtHamiltonian} can be obtained from its previous step using the Woodbury Matrix Identity. Thus the partition function with this ``dreamt" Hamiltonian is given by,
\begin{align}
    Z = \sum_{\sigma} \rm \exp(\frac{\beta}{N}\sum_{\mu \nu i j} \sigma_i \xi^{\mu}_i (1 + \lambda^2 C)^{-1} \xi^{\nu}_j \sigma_j)
\end{align}

\section{Restricted Boltzmann Machines for numerical experiments}
\label{RBMNumericalExpt}
In this section, we look at each of the steps in detail. We follow Ref.~\cite{leonelli2021effective} but make changes for the 3-layered RBM architecture.

\subsection{Data generation}
First, we choose the pattern loading of the system ($\alpha$). We choose two temperatures for sampling with passive dynamics ($T_p$) and with active dynamics ($T_a$). These temperatures and $\alpha$ are chosen such that they are regions where the pattern configurations are the global minima. 

\subsection{Training the RBM}
\label{appsec:trainingRBM}
The RBM architechture consists of N visible nodes corresponding to the N spins of the system, M hidden nodes where M is the number of patterns stored in the original system and N hidden nodes to account for the ``pseudoinverse" structure of the connectivity matrix. The RBM weights are initialized as,
\begin{align}
    W^{\mu}_{i} = \xi^{\mu}_i + z, \ z \sim \mathcal{N}(0,1)
\end{align}
In this section, for convenience, we will denote the temperature at which the RBM samples as $T$ instead of $T_{\rm RBM}$ (and similarly, $\beta$ for $\beta_{\rm RBM}$). It should be noted that $\beta_{\rm RBM}$ is a parameter of the model, not the inverse temperature of the simulations. For the passive case, we know the ground truth that $\beta_{\rm RBM}$ is equal to $\beta$ of the simulations. 

For each sample of the data, the visible nodes are set to the sample configuration,  $\sigma_i = \sigma_i^{data}$. Now give the data, we generate a configuration for the hidden nodes ($z$ and $\phi$). We can use the conditional probabilities for this purpose.
\begin{align}
    &P(z, \phi | \sigma) = \frac{P(z,\phi,\sigma)}{P(\sigma)} = \frac{\exp(-\beta H)}{\int Dz D\phi \exp(-\beta H)} \\
    =& \exp( -\frac{\beta}{2} \left( \phi_i - \frac{i \lambda}{\sqrt{N}}  z_{\mu} W^{\mu}_{i} \right)^2 ) \times \nonumber \\
    & \exp( -\frac{\beta}{2} \left( z_{\mu}  - \frac{W^{\kappa}_i \sigma_i}{\sqrt{N}}A^{-1}_{\kappa \mu} \right) A_{\mu \nu} \left( z_{\nu}  - \frac{W^{\chi}_j \sigma_j}{\sqrt{N}}A^{-1}_{\chi \nu} \right)) \\
    &A_{\mu \nu} = (\delta_{\mu \nu} + \lambda^2 C_{\mu \nu}) \ , \ C = \frac{1}{N}W^{\mu}_i W^{\nu}_i
\end{align}

Now, the configuration of the hidden nodes is generated as,
\begin{align}
    \bold{z} &\sim \mathcal{N}\left( \frac{\bold{W} \bold{\sigma} \bold{A^{-1}}}{\sqrt{N}},T\bold{A^{-1}} \right) \\
    \bold{\phi} &\sim \mathcal{N}\left( \frac{\lambda}{\sqrt{N}} \bold{z} \bold{W}, T\bold{I} \right)
\end{align}

Now, we can regenerate a configuration of the visible nodes using $P(\sigma | z, \phi)$.
\begin{align}
    &P(\sigma|z,\phi) = \frac{P(z,\phi,\sigma)}{P(z,\phi)} = \frac{\exp(-\beta H)}{\sum_{\sigma} \exp(-\beta H)} \\
    &= \frac{1}{1+\rm \exp(-2\beta \frac{W^{\mu}_i}{\sqrt{N}} z_{\mu} \sigma_i)}
\end{align}
Thus, we regenerate the visible nodes as,
\begin{align}
    \sigma'_i = sgn\left( \frac{1}{1+\rm exp(-2\beta \sum_{\mu} W_{\mu i} z_{\mu} )} - r\right), \ r\sim U(0,1)
\end{align}
We again recompute the hidden nodes from these recomputed $\sigma'_i$ as,
\begin{align}
    \bold{z'} &\sim \mathcal{N}\left( \frac{\bold{W} \bold{\sigma'} \bold{A^{-1}}}{\sqrt{N}},T\bold{A^{-1}} \right) \\
    \bold{\phi'} &\sim \mathcal{N}\left( \frac{\lambda}{\sqrt{N}} \bold{z'} \bold{W}, T\bold{I} \right)
\end{align}
The weight matrix and $\lambda$ can now be updated using contrastive divergence. Essentially, what the RBM is trying to do is reproduce the same distribution as the one provided to it from the real world model. If we denote the real data distribution as $Q(\sigma)$ and the model as $P(\sigma;\theta)$ where $\theta$ are the parameters we want to optimize.

\begin{align}
    &D_{KL}(Q||P) = \sum_{\sigma} Q(\sigma) \ln{\frac{Q(\sigma)}{P(\sigma)}} \\
    &\frac{D_{KL}(Q||P}{\partial \theta} = -\sum_{\sigma} Q(\sigma) \partial_{\theta} \ln{P(\sigma;\theta)} \\
    =& -\sum_{\sigma} Q(\sigma) \partial_{\theta} \ln{\frac{e^{-\tilde{H}(\sigma;\theta)}}{\sum_{\sigma'} e^{- \tilde{H}(\sigma'; \theta)}}} \\
    =& -\sum_{\sigma} Q(\sigma) \left[ -\frac{\tilde{H}(\sigma;\theta)}{\partial \theta} + \sum_{\sigma'} \frac{e^{-\tilde{H}(\sigma';\theta)}}{\sum_{\sigma''} e^{-\tilde{H}(\sigma'';\theta)}} \frac{\partial \tilde{H}(\sigma';\theta)}{\partial \theta}\right] \\
    =& \braket{\frac{\partial \tilde{H}(\sigma ;\theta)}{\partial \theta}}_{data} - \braket{\frac{\partial \tilde{H}(\sigma;\theta)}{\partial \theta}}_{model}
\end{align}
The trainable parameters for our system are, the RBM weights, $W$, the dreaming parameter $\lambda$ and the (``effective") inverse temperature $\beta$ and $\tilde{H} = \beta \left(\frac{z_{\mu}^2}{2} + \frac{\phi_i^2}{2} - z_{\mu} \frac{\xi^{\mu}_i}{\sqrt{N}}(\sigma_i + i\lambda \phi_i) \right)$. using this, we have the relaxation equations for the trainable degrees of freedom,
\begin{align}
    \frac{\partial W^{\mu}_i}{\partial t} =& -\frac{\partial D_{KL}(Q||P)}{\partial W^{\mu}_i} \\
    =& \braket{\frac{\partial \tilde{H}(\sigma;\theta)}{\partial \theta}}_{model} - \braket{\frac{\partial \tilde{H}(\sigma;\theta)}{\partial \theta}}_{data} \\
    =& \frac{\beta}{\sqrt{N}} \left( z_{\mu}(\sigma_i+i\lambda \phi_i) - z'_{\mu}(\sigma'_i+i\lambda \phi'_i) \right) \\
    \frac{\partial \lambda}{\partial t} =& i\frac{\beta}{\sqrt{N}} \sum_{\mu,i}W^{\mu}_i( z_{\mu} \phi_i - z'_{\mu} \phi'_i ) \\
    \frac{\partial \beta}{\partial t} =& \frac{1}{\sqrt{N}} \sum_{\mu, i} W^{\mu}_i\left( z_{\mu}(\sigma_i+i\lambda \phi_i) - z'_{\mu}(\sigma'_i+i\lambda \phi'_i) \right) \nonumber\\
    &+ \sum_{\mu} \frac{z_{\mu}^{\prime^2} - z_{\mu}^2}{2} + \sum_{i} \frac{\phi_{i}^{\prime^2} - \phi_{i}^2}{2}
\end{align}
At each step of the update, the columns of the $W$ matrix are regularized such that their norm is $\sqrt{N}$.

\subsection{Different modes of training}
\label{DifferentModesOfTraining}
For data generated using equilibrium dynamics, the patterns, $\xi^{\mu}_i$, the inverse temperature, $\beta$ and $\lambda=0$ are the ground truths of the system whereas for the data generated using active dynamics, only the patterns are the ground truths and one needs to infer the ``effective" temperature and $\lambda$. We train the system using three different modes:
\begin{enumerate}
    \item Learn all $\xi$, $\lambda$ and $\beta$ - If we initialize the parameters at a small distance away from the actual parameters, the RBM gets stuck in a set of parameters which are not the ground truths of the system. We can characterize this well because we know the ground truth $\xi$, $\lambda$ and $\beta$ for the passive simulations exactly. The RBM fails to reach the target $\beta$ and $\lambda$. 
    \item Fix the $\beta$ as the inverse temperature used to generate data in both active and passive simulations. Learn $\xi$ and $\lambda$ - This setting is not ideal for active simulations as we do not know the ``effective" temperature for active simulations.
    \item Fix the weights of the RBM as the patterns, $\xi$. Learn the inverse temperature, $\beta$ and $\lambda$ - This is the ideal setting for comparing active and passive simulations. The patterns, $\xi$ are the ground truths for both passive and active simulations, so constraining the RBM weights using these is a valid thing to do. Since the ``effective" temperature and $\lambda$ for active sims is unknown whereas that for passive sims is known, this mode is ideal for comparing the learning parameter, $\lambda$.
\end{enumerate}

For generating Fig.~\ref{ActiveBetterThanPassive}(b) and (c), we use mode (3) for inferring the parameters. In the next section, we benchmark our RBM inference procedures.

\section{Simulation Details and Checks}
We perform all simulations and inference with $N=50$ spins. The loading capacity $\alpha$ is varied from $0.06$ to $0.1$. The temperature for generating data is varied from $T=0.4$ to $T=0.8$. The degree of activity is varied between $a=0.0$ to $a=0.8$ with $a=0.0$ corresponding to the passive case. The learnable parameters are $\lambda$, $T_{\rm RBM}$ and the patterns $\xi^{\mu}_i (W^{\mu}_i)$.
\subsection{Benchmarking the model with data generated from a Pseudo-inverse Hopfield Hamiltonian}
\label{subsec:pseudobenchmark}
\begin{figure}[ht]
    \centering
    \includegraphics[scale=0.45]{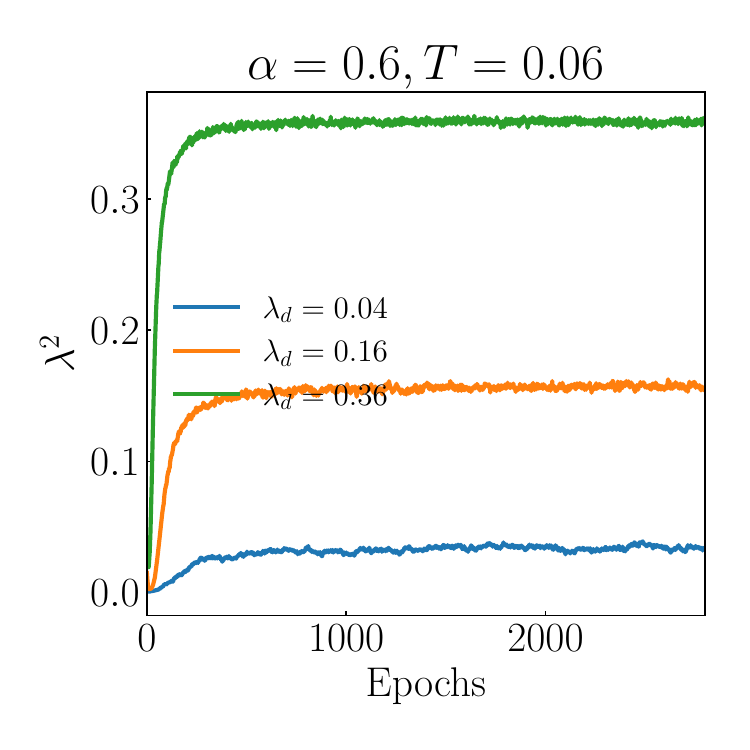}
    \caption{The inferred value of $\lambda$ matches closely with the value used for generating the data. The details are provided in Sec.~\ref{subsec:pseudobenchmark}}
    \label{BenchmarkingPseudo}
\end{figure}

In this benchmarking procedure, $T_{\rm RBM}$ is set to $T$ since the data is generated using passive simulations at a specific temperature, $T$, using the pseudoinverse Hamiltonian and we know exactly that $T_{\rm RBM} = T$. The system infers the $\lambda$ and the patterns $\xi^{\mu}_i$. To be concise, only the inferred $\lambda$ is plotted. The three plots in Fig.~\ref{BenchmarkingPseudo}, are for data generated using three different values of $\lambda$ as it is varied between $0.2$ to $0.6$. This shows that our system is capable of inferring the extent of dreaming, $\lambda$, accurately.

\subsection{Varying Initial Conditions}
\label{subsec:varyInitialT}
\begin{figure}[ht]
    \centering
    \includegraphics[scale=0.35]{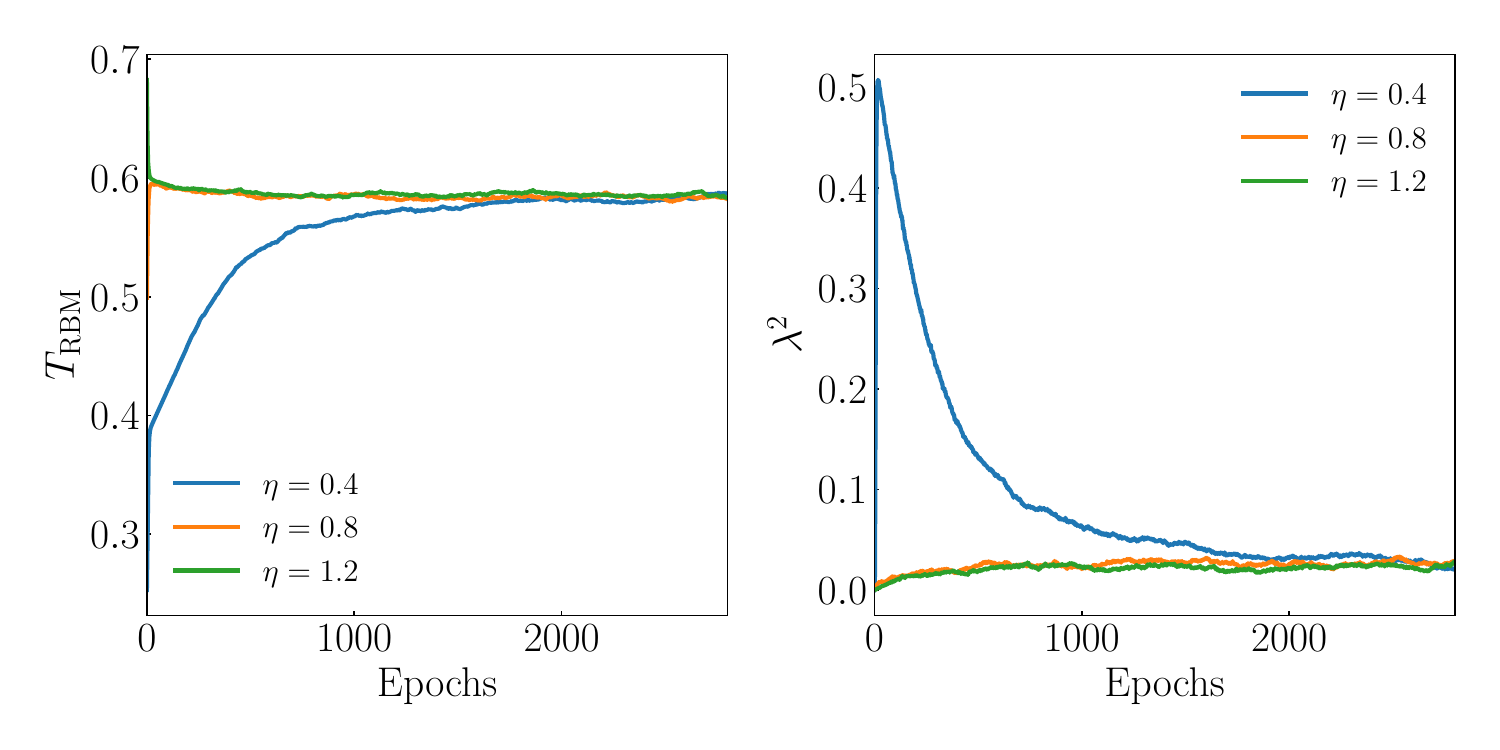}
    \caption{The system learns the $T_{\rm RBM}$ from data when initialized at $0.4T$, $0.8T$ and $1.2T$. Also, the $\lambda$ learnt is very close to 0, as it should be. The details about the simulations are provided in Sec.~\ref{subsec:varyInitialT}.}
    \label{VaryingTinit}
\end{figure}

For this set of simulations, we fix the weights of the RBM to be equal to the patterns, $W^{\mu}_i = \xi^{\mu}_i$ and learn only the $\lambda$ and $T_{\rm RBM}$. The initial conditions for learning $T_{\rm RBM}$ are varied from $T_{RBM}^{init}=0.4T$ to $1.2T$. The data for plots in Fig.~\ref{VaryingTinit} are  generated using $\alpha=0.06$, $T=0.06$, $a=0$ i.e. these are passive simulations for which we know that $T_{\rm RBM}=T$ and $\lambda=0$. This shows that our system is robust to initial conditions of $T_{\rm RBM}$ and infers it with good accuracy while inferring the corresponding $\lambda$ as well.

\subsection{Vary the $T_{RBM}$ and run various active and passive simulations}
\label{subsec:VaryTRBM}
\begin{figure}[ht]
    \centering
    \includegraphics[scale=0.45]{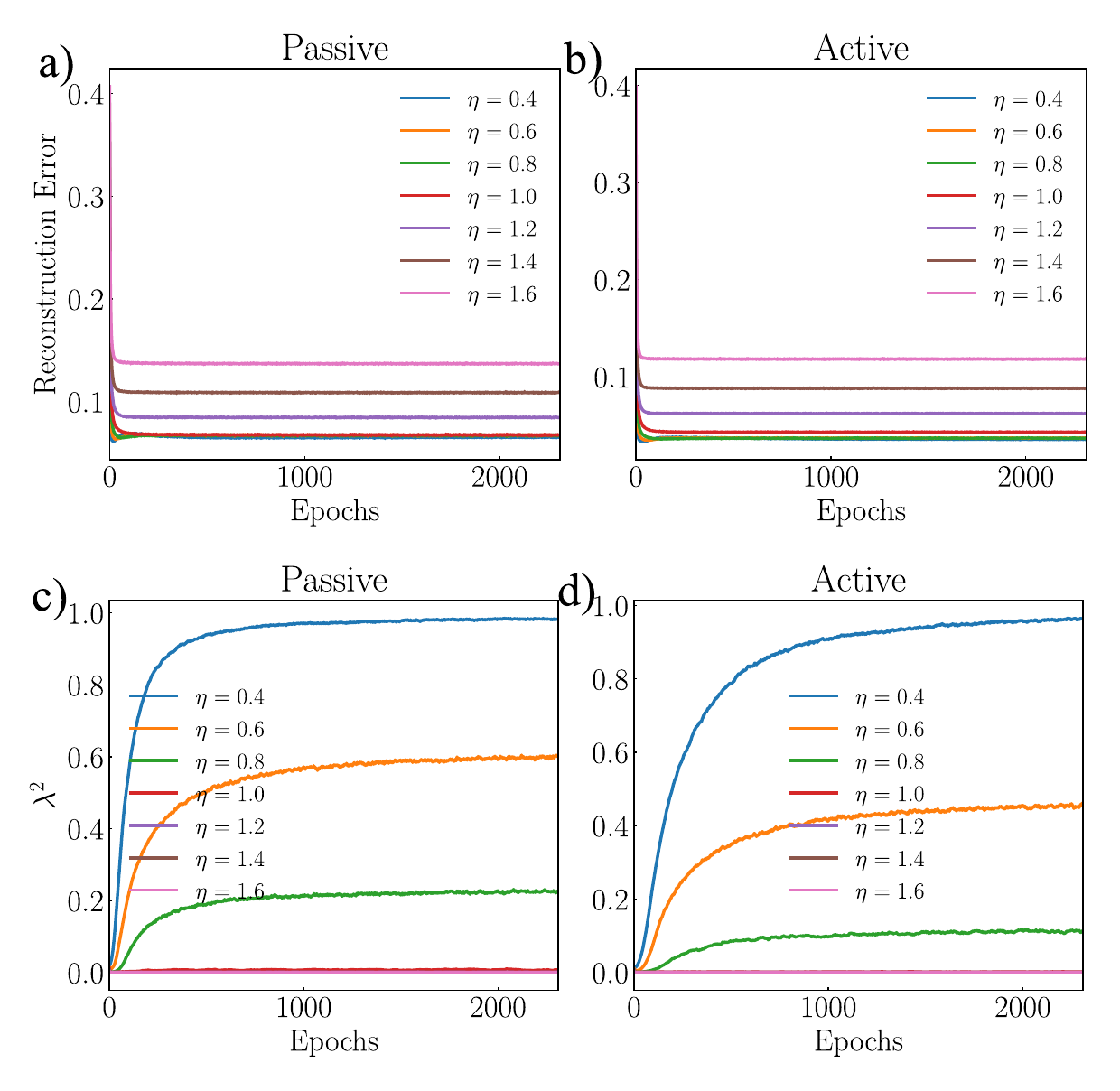}
    \caption{The $\lambda$ corresponding to the saturation of RCE is non-zero for active case whereas it is $0$ for passive case. The details of the simulations are provided in Sec.~\ref{subsec:VaryTRBM}.}
    \label{RCE_lambda_Vary_T}
\end{figure}

For the set of inference in Fig.~\ref{RCE_lambda_Vary_T}, we fix $T_{\rm RBM}$ and learn $\lambda$ and the weights $W^{\mu}_i$. $T_{\rm RBM}$ is set to $\eta T$, where $\eta$ varies from $1.6$ to $0.4$ in steps of $0.2$ and $T$ is the temperature at which the data has been generated and is set to $T=0.4$. Both passive and active data are generated with pattern loading, $\alpha=0.06$. For active data, $a=0.4$. The reconstruction error (RCE) is plotted in panels (a) and (b) and the corresponding $\lambda$ values are plotted in (c) and (d). As $T_{\rm RBM}$ is decreased from $1.6T$ to $0.4T$, the steady state RCE value goes down but saturates at some point, i.e. does not go down any further. This value of $T_{\rm RBM}$ is closest to the actual temperature used to generate the data. For the passive data, RCE saturates at $T_{\rm RBM} = 1.0T$ and the corresponding $\lambda$ value is close to $0$. This is expected. For active simulations, RCE saturates at $T_{\rm RBM} = 0.8T$ indicating that $0.8T$ is closest to the notional ``temperature" for active case and the corresponding $\lambda$ is non-zero. Since there does not exist a notion of temperature for active simulations, this is a good way to understand that when we try to distribution generated by active simulations with our \textit{ansatz}, the ``effective" temperature is different from the ``temperature" parameter used to carry out the simulations. Thus, in order to compare the passive and active data on an equal footing we must learn $T_{\rm RBM}$ along with other parameters. Fixing $T_{\rm RBM}$ might lead to incorrect inference in the active case.

\subsection{Reconstruction Error for Fig.~\ref{ActiveBetterThanPassive} of Main text}
\begin{figure}[ht]
    \centering
    \includegraphics[scale=0.35]{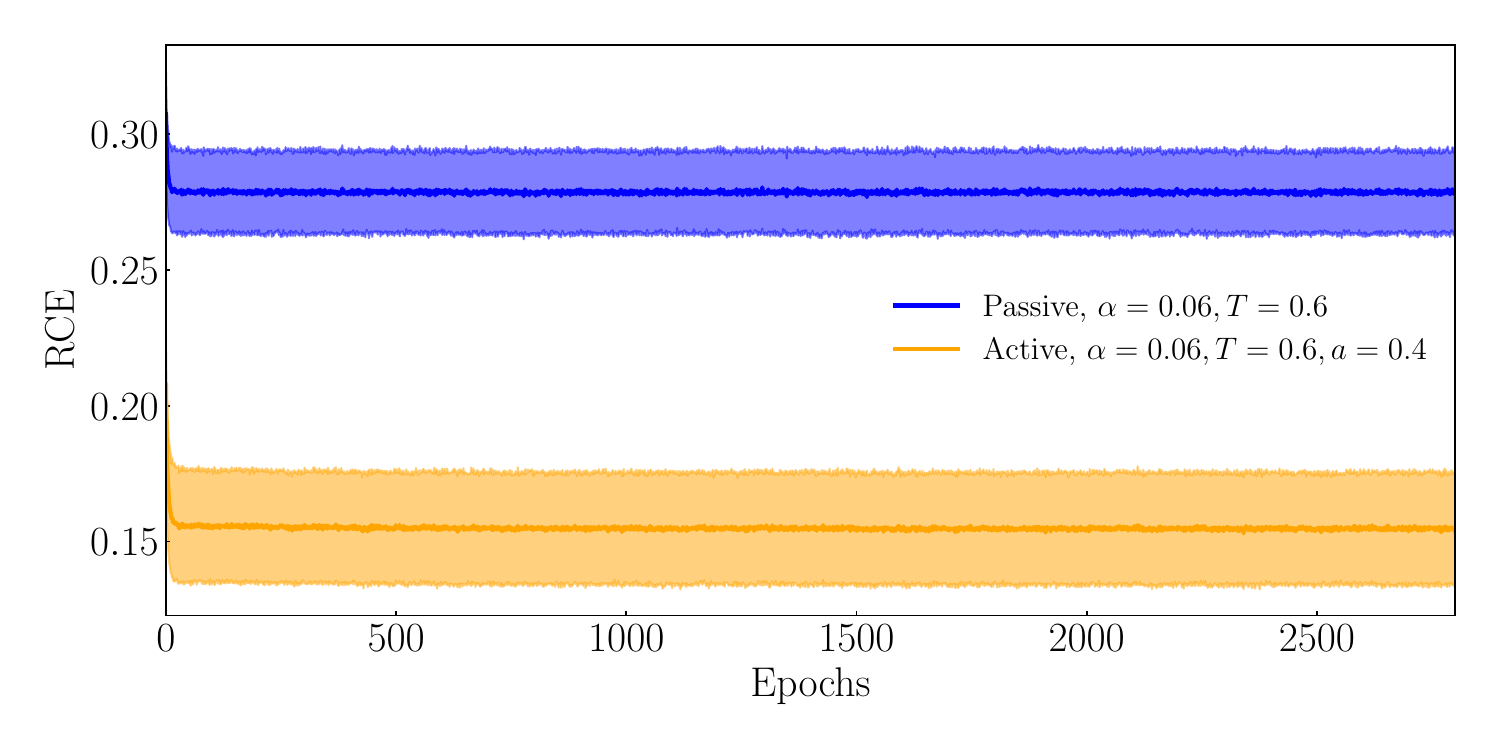}
    \caption{Reconstruction error as a function of epochs for $\alpha=0.06$, $T=0.6$ for both active and passive cases.}
    \label{fig:RCEforMainFig}
\end{figure}

\bibliographystyle{unsrt}
\bibliography{references}

\begin{thebibliography}{10}

\bibitem{marullo2020boltzmann}
Chiara Marullo and Elena Agliari.
\newblock Boltzmann machines as generalized hopfield networks: a review of recent results and outlooks.
\newblock {\em Entropy}, 23(1):34, 2020.

\bibitem{le2008representational}
Nicolas Le~Roux and Yoshua Bengio.
\newblock Representational power of restricted boltzmann machines and deep belief networks.
\newblock {\em Neural computation}, 20(6):1631--1649, 2008.

\bibitem{shimagaki2019selection}
Kai Shimagaki and Martin Weigt.
\newblock Selection of sequence motifs and generative hopfield-potts models for protein families.
\newblock {\em Physical Review E}, 100(3):032128, 2019.

\bibitem{schnaack2022learning}
Oskar~H Schnaack, Luca Peliti, and Armita Nourmohammad.
\newblock Learning and organization of memory for evolving patterns.
\newblock {\em Physical Review X}, 12(2):021063, 2022.

\bibitem{brennan1990olfactory}
Peter Brennan, Hideto Kaba, and Eric~B Keverne.
\newblock Olfactory recognition: a simple memory system.
\newblock {\em Science}, 250(4985):1223--1226, 1990.

\bibitem{haberly1989olfactory}
Lewis~B Haberly and James~M Bower.
\newblock Olfactory cortex: model circuit for study of associative memory?
\newblock {\em Trends in neurosciences}, 12(7):258--264, 1989.

\bibitem{wilson2004plasticity}
DA~Wilson, AR~Best, and RM~Sullivan.
\newblock Plasticity in the olfactory system: lessons for the neurobiology of memory.
\newblock {\em The Neuroscientist}, 10(6):513--524, 2004.

\bibitem{barton2015scaling}
John~P Barton, Mehran Kardar, and Arup~K Chakraborty.
\newblock Scaling laws describe memories of host--pathogen riposte in the hiv population.
\newblock {\em Proceedings of the National Academy of Sciences}, 112(7):1965--1970, 2015.

\bibitem{cocco2011adaptive}
Simona Cocco and R{\'e}mi Monasson.
\newblock Adaptive cluster expansion for inferring boltzmann machines with noisy data.
\newblock {\em Physical review letters}, 106(9):090601, 2011.

\bibitem{palmer2015predictive}
Stephanie~E Palmer, Olivier Marre, Michael~J Berry, and William Bialek.
\newblock Predictive information in a sensory population.
\newblock {\em Proceedings of the National Academy of Sciences}, 112(22):6908--6913, 2015.

\bibitem{krotov2023new}
Dmitry Krotov.
\newblock A new frontier for hopfield networks.
\newblock {\em Nature Reviews Physics}, 5(7):366--367, 2023.

\bibitem{boukacem2024waddington}
Nacer~Eddine Boukacem, Allen Leary, Robin Th{\'e}riault, Felix Gottlieb, Madhav Mani, and Paul Fran{\c{c}}ois.
\newblock Waddington landscape for prototype learning in generalized hopfield networks.
\newblock {\em Physical Review Research}, 6(3):033098, 2024.

\bibitem{hopfield1983unlearning}
John~J Hopfield, David~I Feinstein, and Richard~G Palmer.
\newblock ‘unlearning’has a stabilizing effect in collective memories.
\newblock {\em Nature}, 304(5922):158--159, 1983.

\bibitem{derrida1987exactly}
Bernard Derrida, Elizabeth Gardner, and Anne Zippelius.
\newblock An exactly solvable asymmetric neural network model.
\newblock {\em Europhysics Letters}, 4(2):167, 1987.

\bibitem{hopfield1982neural}
John~J Hopfield.
\newblock Neural networks and physical systems with emergent collective computational abilities.
\newblock {\em Proceedings of the national academy of sciences}, 79(8):2554--2558, 1982.

\bibitem{amit1989modeling}
Daniel~J Amit and Daniel~J Amit.
\newblock {\em Modeling brain function: The world of attractor neural networks}.
\newblock Cambridge university press, 1989.

\bibitem{amit1985storing}
Daniel~J Amit, Hanoch Gutfreund, and Haim Sompolinsky.
\newblock Storing infinite numbers of patterns in a spin-glass model of neural networks.
\newblock {\em Physical Review Letters}, 55(14):1530, 1985.

\bibitem{dotsenko1991statistical}
VS~Dotsenko, ND~Yarunin, and EA~Dorotheyev.
\newblock Statistical mechanics of hopfield-like neural networks with modified interactions.
\newblock {\em Journal of Physics A: Mathematical and General}, 24(10):2419, 1991.

\bibitem{fodor2016far}
{\'E}tienne Fodor, Cesare Nardini, Michael~E Cates, Julien Tailleur, Paolo Visco, and Fr{\'e}d{\'e}ric Van~Wijland.
\newblock How far from equilibrium is active matter?
\newblock {\em Physical review letters}, 117(3):038103, 2016.

\bibitem{hopfield1984neurons}
John~J Hopfield.
\newblock Neurons with graded response have collective computational properties like those of two-state neurons.
\newblock {\em Proceedings of the national academy of sciences}, 81(10):3088--3092, 1984.

\bibitem{behera2023enhanced}
Agnish~Kumar Behera, Madan Rao, Srikanth Sastry, and Suriyanarayanan Vaikuntanathan.
\newblock Enhanced associative memory, classification, and learning with active dynamics.
\newblock {\em Physical Review X}, 13(4):041043, 2023.

\bibitem{agliari2019dreaming}
Elena Agliari, Francesco Alemanno, Adriano Barra, and Alberto Fachechi.
\newblock Dreaming neural networks: rigorous results.
\newblock {\em Journal of Statistical Mechanics: Theory and Experiment}, 2019(8):083503, 2019.

\bibitem{leonelli2021effective}
Francesca~Elisa Leonelli, Elena Agliari, Linda Albanese, and Adriano Barra.
\newblock On the effective initialisation for restricted boltzmann machines via duality with hopfield model.
\newblock {\em Neural Networks}, 143:314--326, 2021.

\bibitem{crair1989non}
Michael Crair and William Bialek.
\newblock Non-boltzmann dynamics in networks of spiking neurons.
\newblock {\em Advances in neural information processing systems}, 2, 1989.

\bibitem{stuart2016dendrites}
Greg Stuart, Nelson Spruston, H{\"a}usser Michael, et~al.
\newblock {\em Dendrites}.
\newblock Oxford University Press, 2016.

\bibitem{dayan2005theoretical}
Peter Dayan and Laurence~F Abbott.
\newblock {\em Theoretical neuroscience: computational and mathematical modeling of neural systems}.
\newblock MIT press, 2005.

\bibitem{zheng2012predictive}
Weihua Zheng, Nicholas~P Schafer, Aram Davtyan, Garegin~A Papoian, and Peter~G Wolynes.
\newblock Predictive energy landscapes for protein--protein association.
\newblock {\em Proceedings of the National Academy of Sciences}, 109(47):19244--19249, 2012.

\bibitem{beissinger1998chaperones}
Martina Beissinger and J~Buchner.
\newblock How chaperones fold proteins.
\newblock {\em Biological chemistry}, 379(3):245--259, 1998.

\bibitem{goloubinoff2018chaperones}
Pierre Goloubinoff, Alberto~S Sassi, Bruno Fauvet, Alessandro Barducci, and Paolo De~Los~Rios.
\newblock Chaperones convert the energy from atp into the nonequilibrium stabilization of native proteins.
\newblock {\em Nature chemical biology}, 14(4):388--395, 2018.

\bibitem{saibil2013chaperone}
Helen Saibil.
\newblock Chaperone machines for protein folding, unfolding and disaggregation.
\newblock {\em Nature reviews Molecular cell biology}, 14(10):630--642, 2013.

\end{thebibliography}

\end{document}